\documentclass[12pt]{iopart}
\usepackage{graphicx}
\usepackage{iopams}
\usepackage{cite}
\newcommand{\avg}[1]{\langle #1 \rangle}

\newcommand{\eqref}[1]{\eref{#1}}
\newcommand{\tfrac}[2]{\case{#1}{#2}}

\begin{document}

\title[Oscillatory states, standing waves, and additional synchronized clusters]{Oscillatory states, standing waves, and additional synchronized clusters in networks of second-order oscillators: uncovering the role of inertia}

\author{Jian Gao$^1$, Konstantinos Efstathiou$^2$}
\address{$^1$ Bernoulli Institute for Mathematics, Computer Science, and Artificial Intelligence, University of Groningen, P.O. Box 407, 9700 AK, Groningen, The Netherlands}
\address{$^2$ Division of Natural and Applied Sciences and Zu Chongzhi Center for Mathematics and Computational Science, Duke Kunshan University, No.\ 8 Duke Avenue, Kunshan 215316, China}
\ead{k.efstathiou@dukekunshan.edu.cn}

\begin{abstract}
We discuss the appearance of oscillatory and standing wave states in second-order oscillator networks showing that it is a special case of a more general mechanism involving secondary synchronized clusters induced by inertia. Using a time-periodic mean-field ansatz, we find a bistable mechanism involving a stable fixed point and an invariant curve of an appropriate Poincar\'e map. The bistability and the devil's staircase associated to the rotation number on the invariant curve provide an explanation for the appearance of the secondary synchronized clusters. The effect of inertias in the self-organization process is analyzed through a simplified model. This shows that the effect of giant synchronized clusters on the other oscillators is weakened by inertias, thus leading to secondary synchronized clusters during the transition process to synchronization. 
\end{abstract}


\section{Introduction}
\label{sec/introduction}

Synchronization of coupled dynamical units is a prevalent phenomenon in nature \cite{Arenas2008,Acebron2005kuramoto} and
many mathematical models have been used in its study.
Among them, coupled Kuramoto oscillators is one of the most popular models \cite{Kuramoto1987,Rodrigues2016}.
Since 1991, second-order Kuramoto oscillators where frequency adaptations (inertias) are added to the Kuramoto model have been proposed to describe the dynamics of three tropical Asian firefly species \cite{Ermentrout1991}.
Several applications of this model have been found, for Josephson junction arrays \cite{Levi1978,Watanabe1994,Trees2005}, goods markets \cite{Ikeda2012}, dendritic neurons \cite{Sakyte2011}, and power grids \cite{Filatrella2008,Rohden2012,Rohden2014,Lozano2012,Witthaut2012,Menck2013,Hellmann2016,Kim2015,Gambuzza2017,Dorfler2013c,Grzybowski2016,Maizi2016,Manik2016a,Pinto2016,Rohden2017,Witthaut2016}.

In this paper we consider a model of coupled second-order oscillators, where the dynamics is given by
\begin{equation}\label{eq_dynamics_original}
  m \ddot{\theta}_i + D\dot{\theta}_i
  = \Omega_i
  + \frac{K}{N}\sum_{j=1}^{N}\sin(\theta_j-\theta_i),
  \quad i=1,\dots,N.
\end{equation}
Here $m$ is the inertia and $D$ the damping coefficient for all oscillators, $N$ is the number of oscillators and $K$ is the coupling strength.
The natural frequencies $\Omega_i$ are randomly chosen from a distribution $g(\Omega)$.
The state of the $i$-th oscillator is described by its phase $\theta_i \in \mathbb{S} = \mathbb{R}/2\pi\mathbb{Z}$. 
The collective state of the oscillators is described by the \emph{order parameter}
\[ r e^{i\phi} = \frac{1}{N} \sum_{j=1}^{N}e^{i\theta_j},\]
where $r$ measures the phase coherence, and $\phi$ represents a collective phase.
If all the oscillators move in a single tight cluster we have $r \approxeq 1$.
On the contrary, if the oscillators move incoherently, scattered around the circle, we have $r \approxeq 0$.

When $m=0$, the second-order oscillators become Kuramoto oscillators.
Ku\-ra\-mo\-to oscillators with a symmetric unimodal distribution $g(\Omega)$ have a continuous synchronization transition with the increase of $K$ from $r \approxeq 0$ to $r\approxeq 1$ \cite{Kuramoto1987}. 
In the presence of inertias, the dynamics of oscillators becomes much more complicated.
In particular, with the increase of $m$, several new features manifest in second-order oscillators, such as hysteresis \cite{Olmi2014}, change of the type of phase transitions \cite{Tanaka1997,Tanaka1997a,Acebron2000synchronization,Barre2016}, and finally oscillatory states with periodic oscillations of the order parameter \cite{Olmi2014}. 
Such oscillatory states are not only found in systems with unimodal distributions of $\Omega$ \cite{Tanaka1997a}, but also with bimodal distributions \cite{Olmi2016} and in complex networks \cite{Olmi2014}.
With the help of the self-consistent method, the dynamics of hysteresis and discontinuous transitions have been recently analyzed in \cite{Gao2018}.
For oscillatory states, Olmi \textit{et al} \cite{Olmi2014} have related the oscillation of the order parameter to the appearance of secondary synchronized clusters using numerical simulations.
However, the dynamics of this oscillatory state and the appearance of additional synchronized clusters is still not well understood. 

\section{Additional synchronized clusters}
\label{sec/numerics}

To provide a more refined description of collective states (compared to the global description provided by the order parameter) we use the mean frequency $\langle\dot{\theta}_j\rangle$ of each oscillator. 
Two oscillators are \emph{synchronized} (or \emph{frequency locked}) if they have the same mean frequency.
A group of oscillators with the same value of mean frequency forms a \emph{synchronized cluster}.
States without any synchronized cluster are steady states with $r = 0$, called \emph{incoherent states}.
States with only one synchronized cluster are \emph{(partial) synchronization states}.
Their order parameters have a constant modulus, $r(t) = r$ and a uniformly rotating phase $\phi=\Omega^r t+\Psi$.
When all the oscillators have the same phase, we have the \emph{complete synchronization state} with $r=1$. If there are more than one synchronized clusters, the order parameter may have a time-dependent modulus, such as in \emph{standing waves} and \emph{oscillating states}.

To explore oscillatory states, we have numerically calculated the dynamics of a network with $N=10000$ oscillators, following Eq.~\eqref{eq_dynamics_original}.
The integration was done using the fourth order Runge-Kutta method with fixed-size time-step $dt = 10^{-3}$.
The natural frequency $\Omega_i$ for each oscillator is chosen randomly from a distribution that is either Gaussian or a double Gaussian.
To describe the upper and lower branches of hysteresis loops, we consider forward and backward processes.
In the forward process the initial states of the oscillators are randomly chosen as $\theta(0) \in [0,2\pi]$, $\dot{\theta}(0) \in [0,1]$ (incoherent state) and then the coupling strength is gradually increased with step $dK = 0.01$.
At each step, the initial states of all the oscillators are the final states in the previous step.
After a transient period $t_0 = 100$, we calculate the order parameter $r$ and mean frequencies $\avg{\dot{\theta}_i}$ over a measurement period $\Delta t=10$ and then move to the next step increasing $K$ by $dK$.
In the backward process, the initial states of the oscillators are randomly chosen as $\theta(0)\in [0,0.02\pi]$, $\dot{\theta}(0)\in[0,1]$ (synchronized state) and the previously described procedure is followed with the value of $K$ decreasing at each step by $dK = -0.01$.

\begin{figure}[tbp]
\centering
\includegraphics[width=0.48\columnwidth]{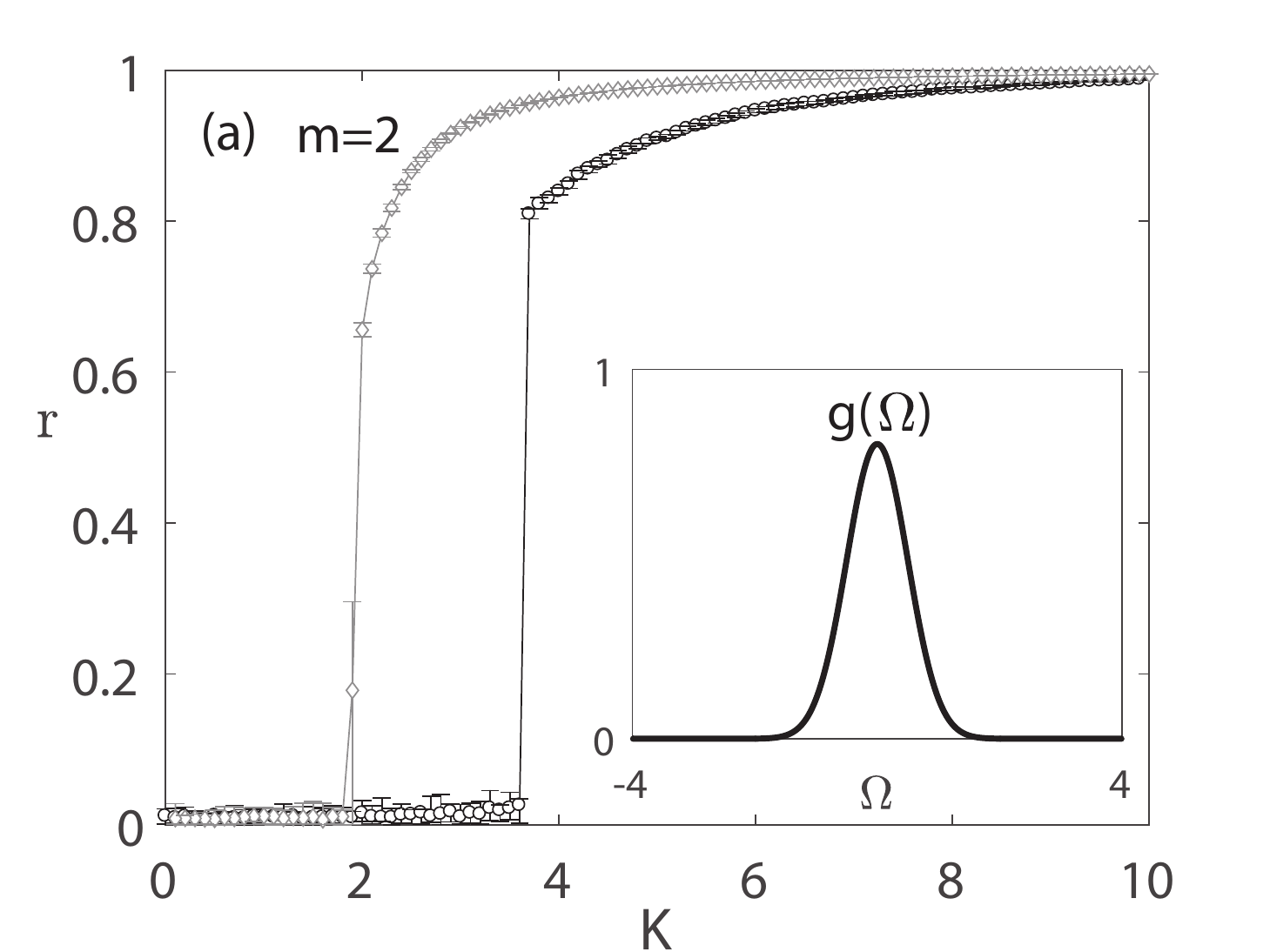}
\includegraphics[width=0.48\columnwidth]{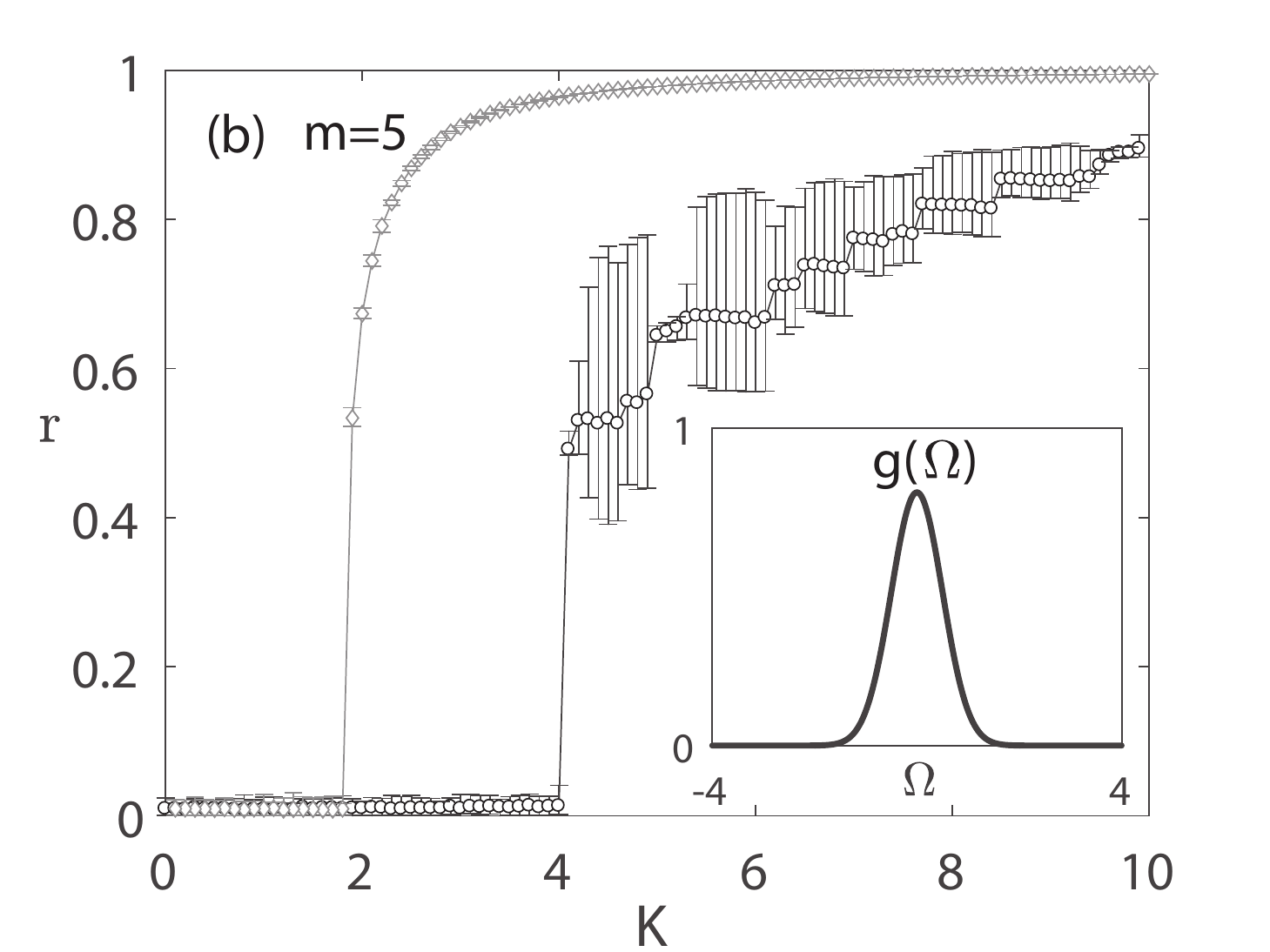}
\includegraphics[width=0.48\columnwidth]{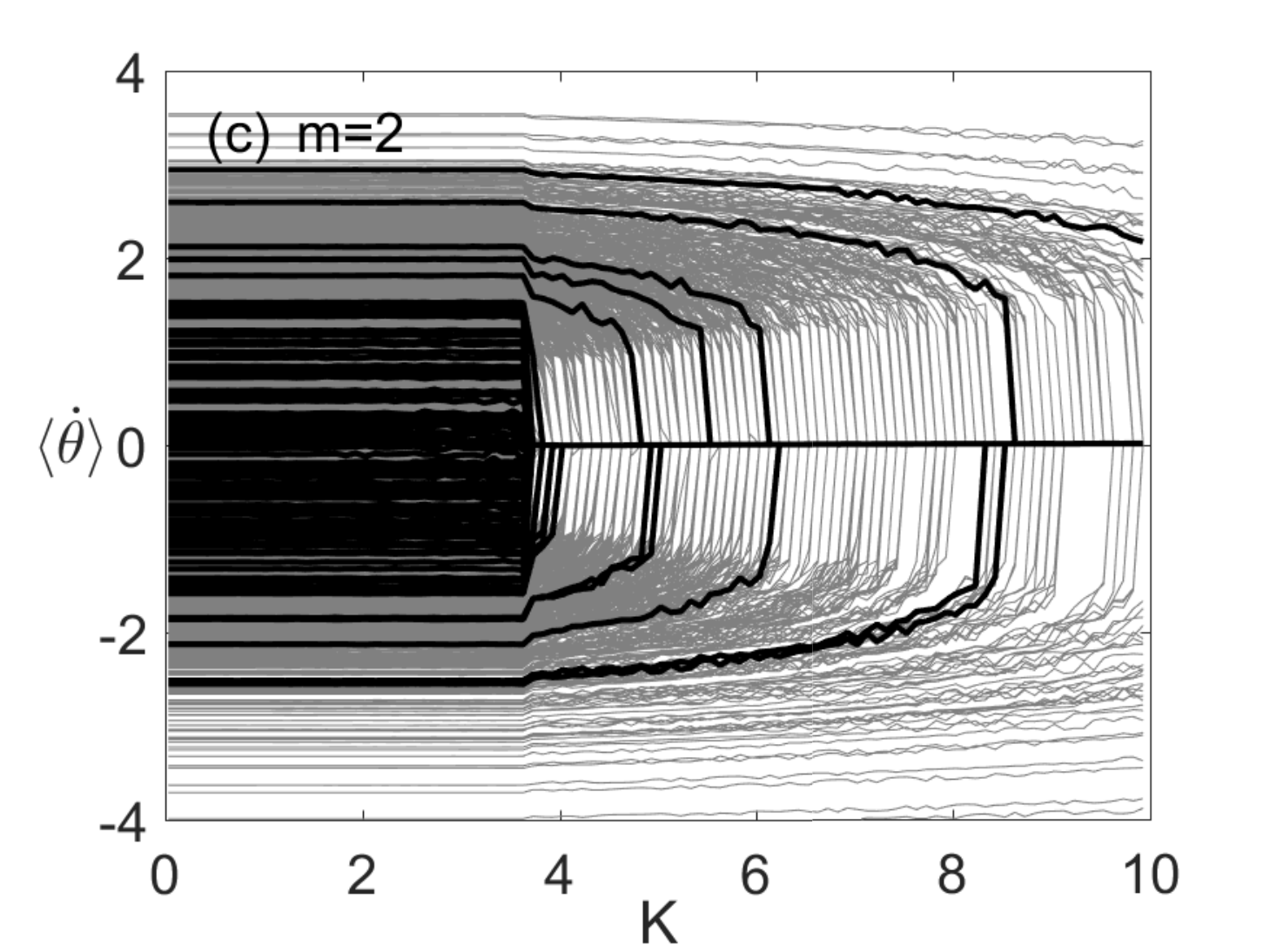}
\includegraphics[width=0.48\columnwidth]{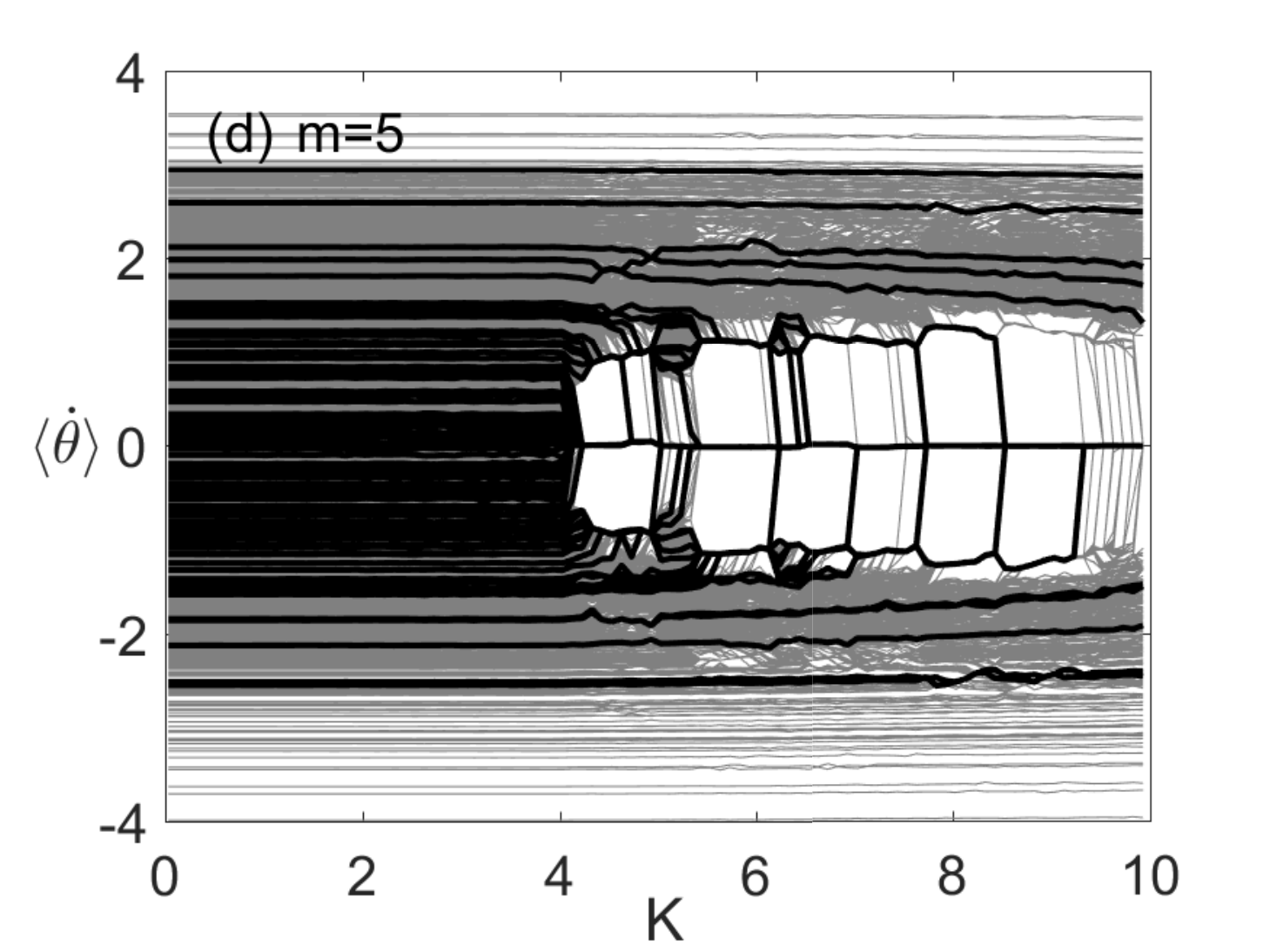}
\includegraphics[width=0.48\columnwidth]{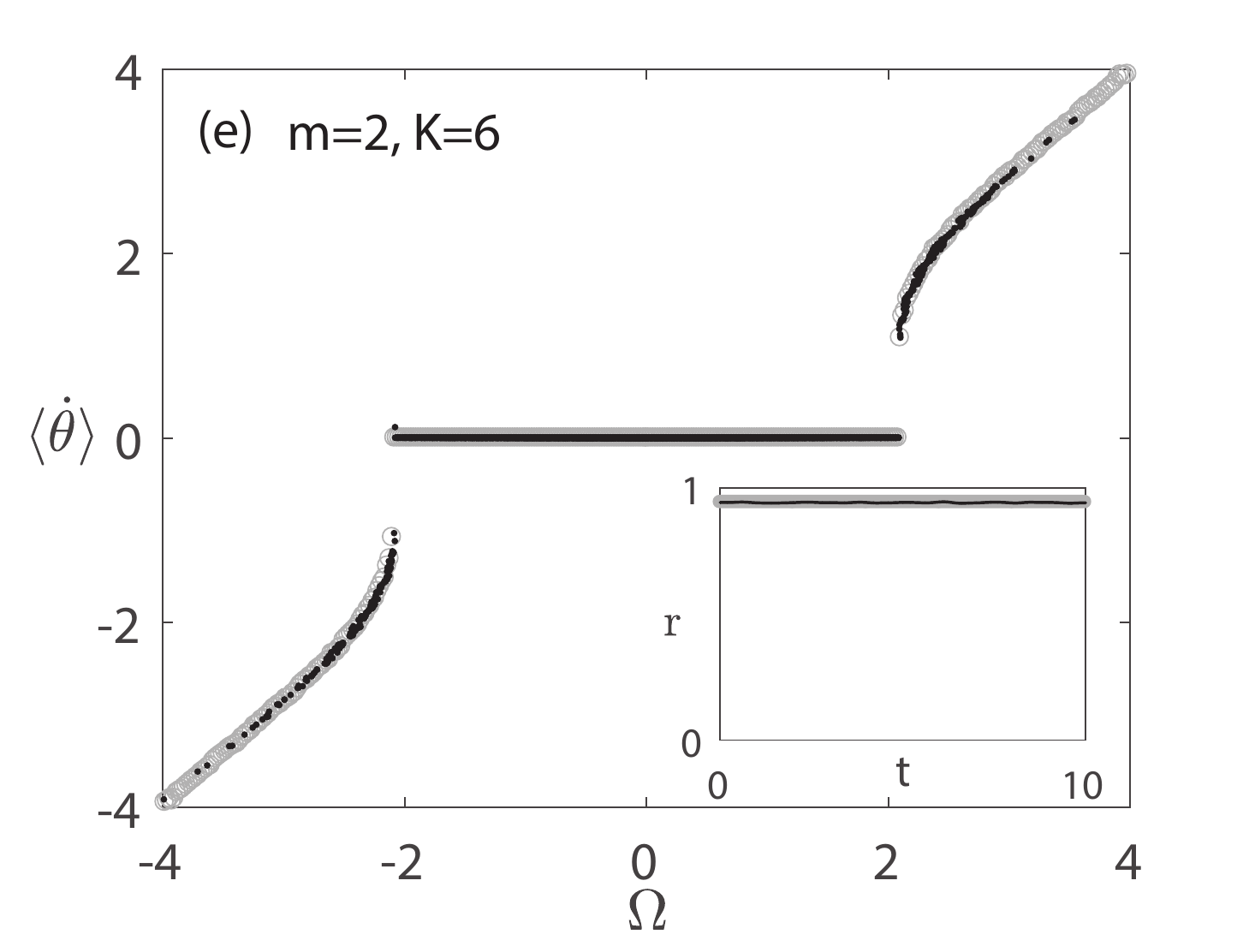}
\includegraphics[width=0.48\columnwidth]{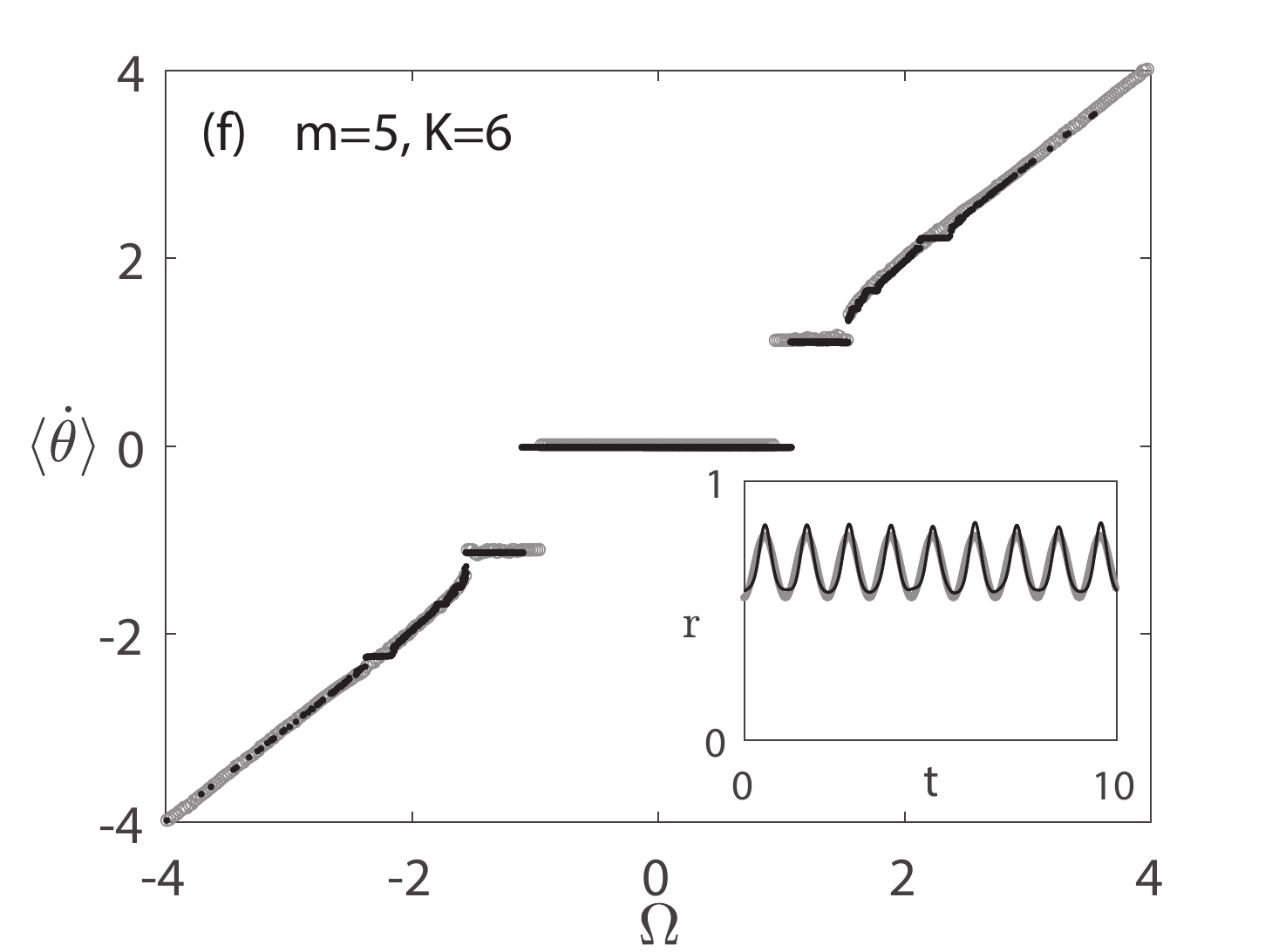}
\caption{Backward and forward processes for $N=10000$ oscillators with (left column) $m=2$ and (right column) $m=5$ for oscillators with Gaussian distributed natural frequencies, Eq.~\eqref{eq_unimodal}.
The black circle (gray diamond) in panels (a,b) are the numerical results in the forward (backward) process showing the evolution of $r$ with decreasing (increasing) $K$.
the error bars show the minimum and maximum values of $r$ for each $K$; oscillatory states correspond to large error bars.
The evolution of the mean frequency $\avg{\dot\theta}$ for the forward process (increasing $K$) is shown in panels (c,d).
The dependence of $\avg{\dot\theta}$ on $\Omega$ is shown panels (e,f) for two typical states with different inertias and same coupling strength $K=6$.
}
\label{fig_unimodal}
\end{figure}

\begin{figure}[tbp]
\centering
\includegraphics[width=0.48\columnwidth]{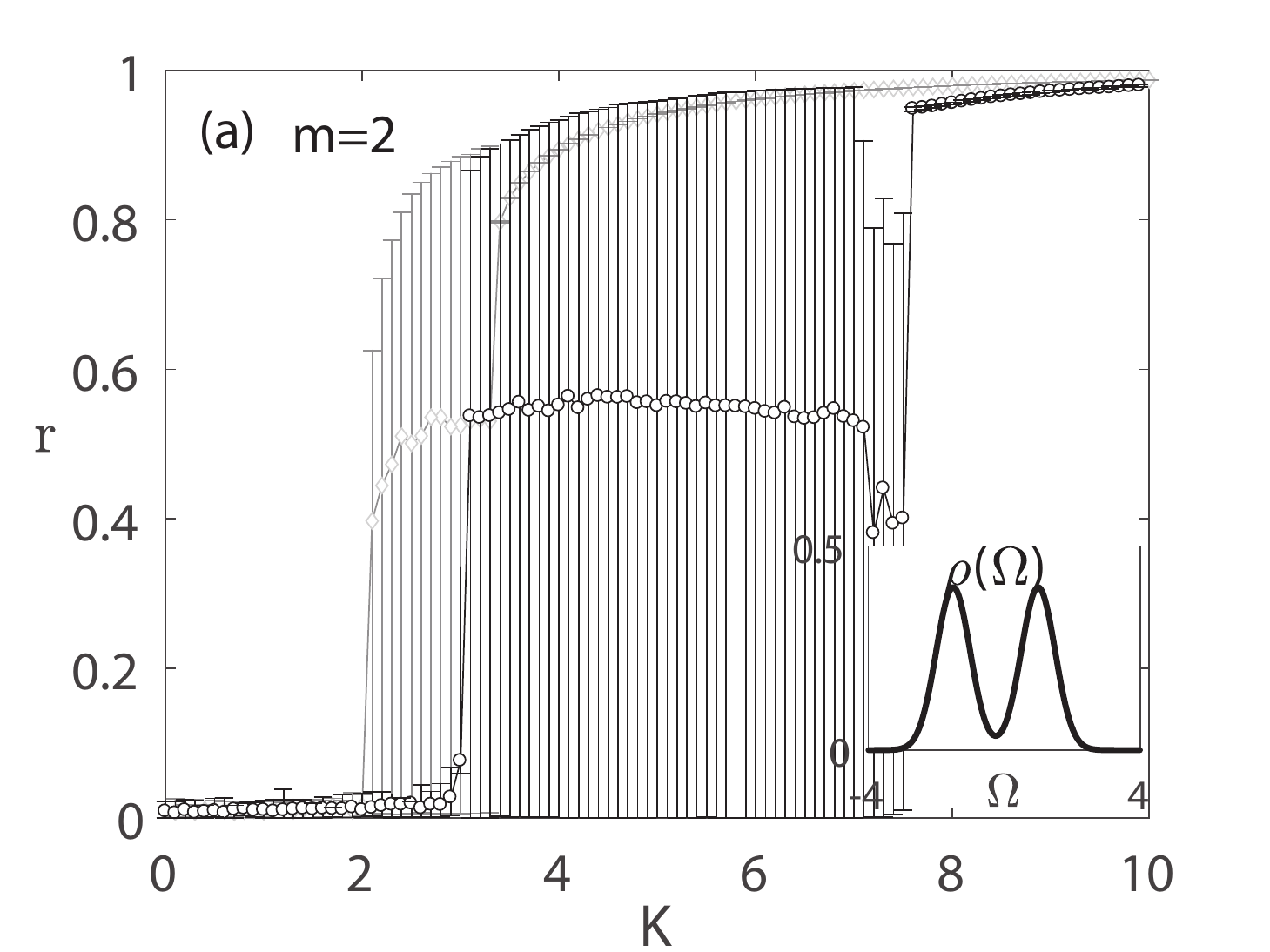}
\includegraphics[width=0.48\columnwidth]{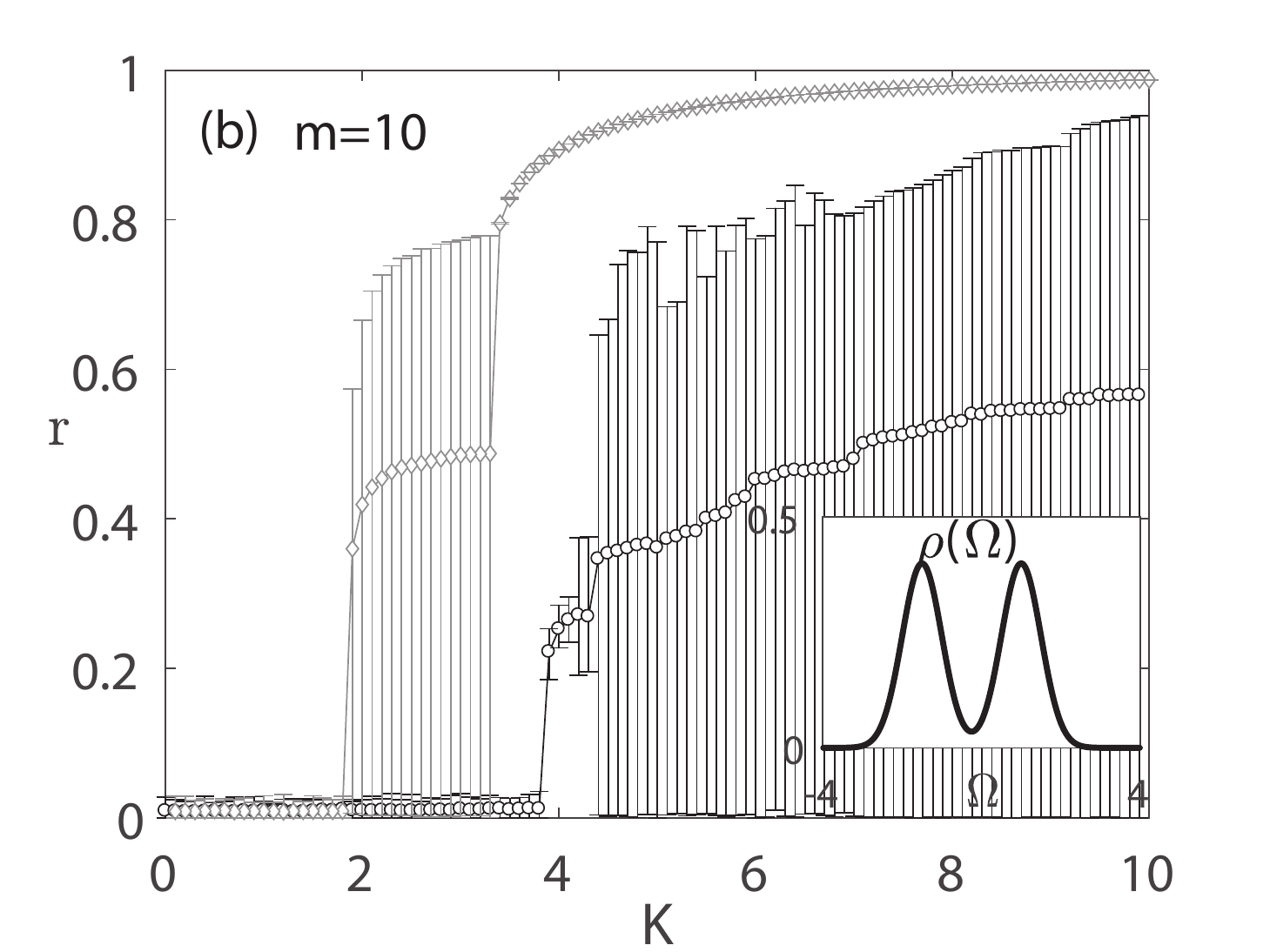}
\includegraphics[width=0.48\columnwidth]{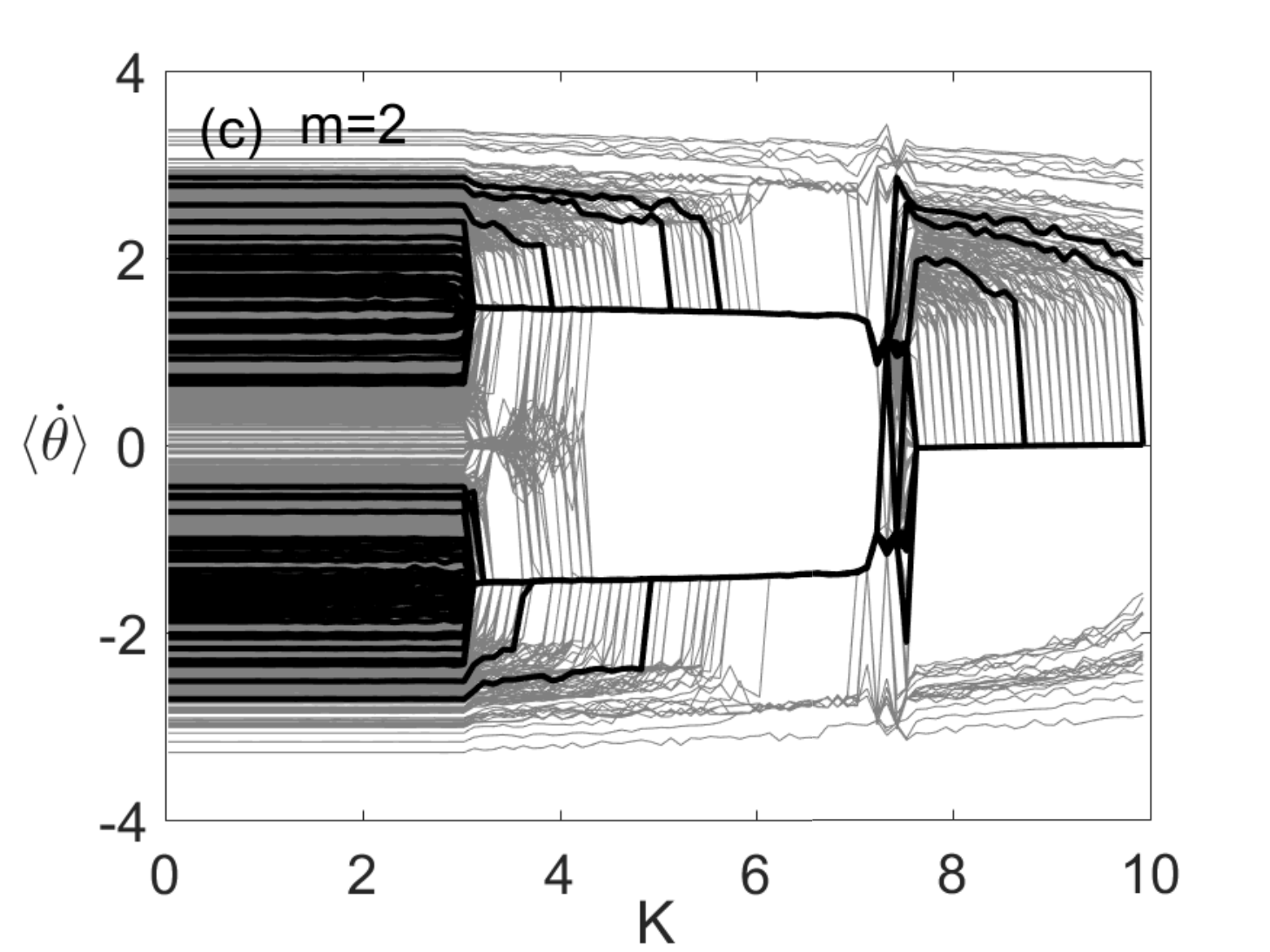}
\includegraphics[width=0.48\columnwidth]{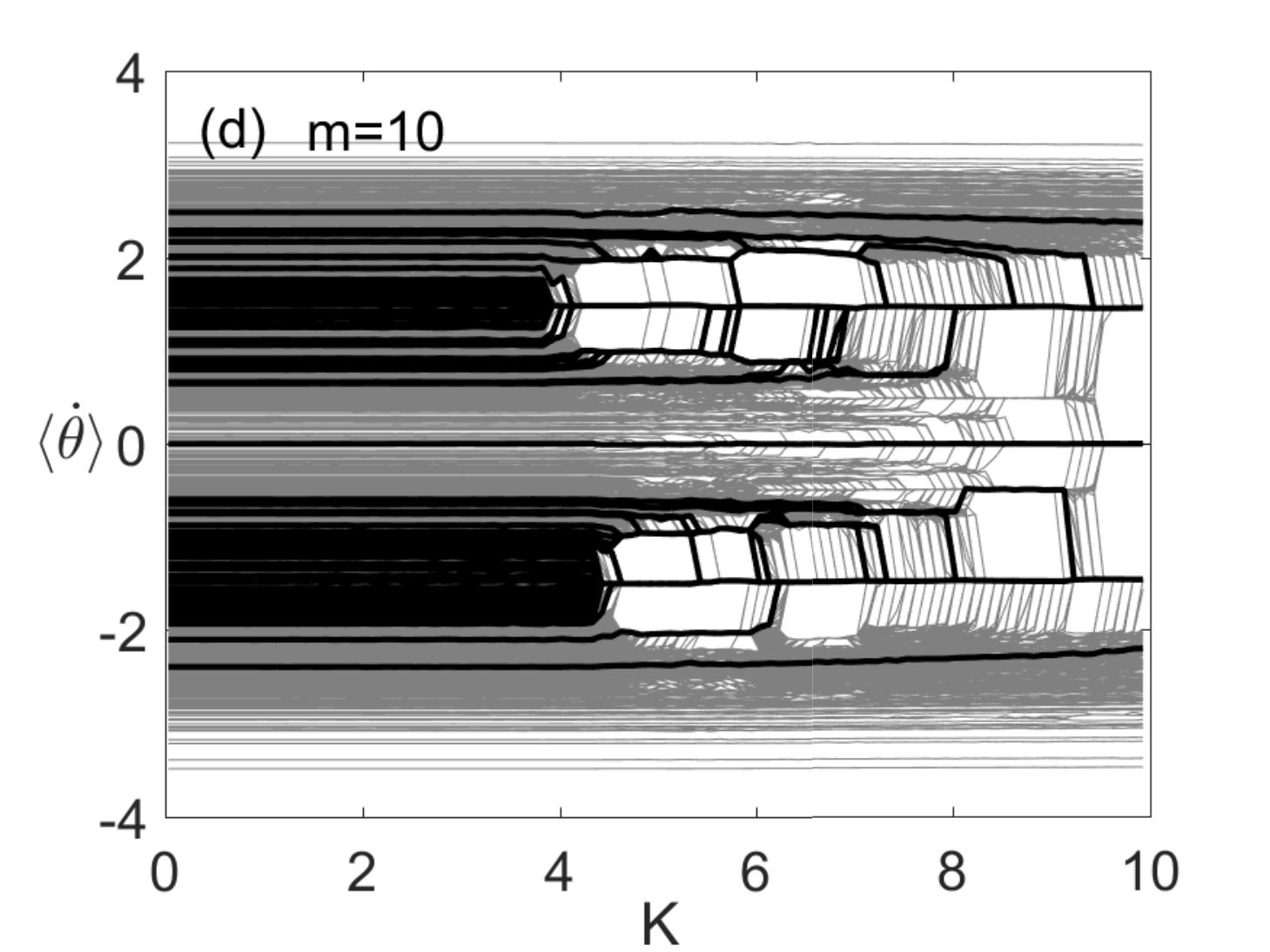}
\includegraphics[width=0.48\columnwidth]{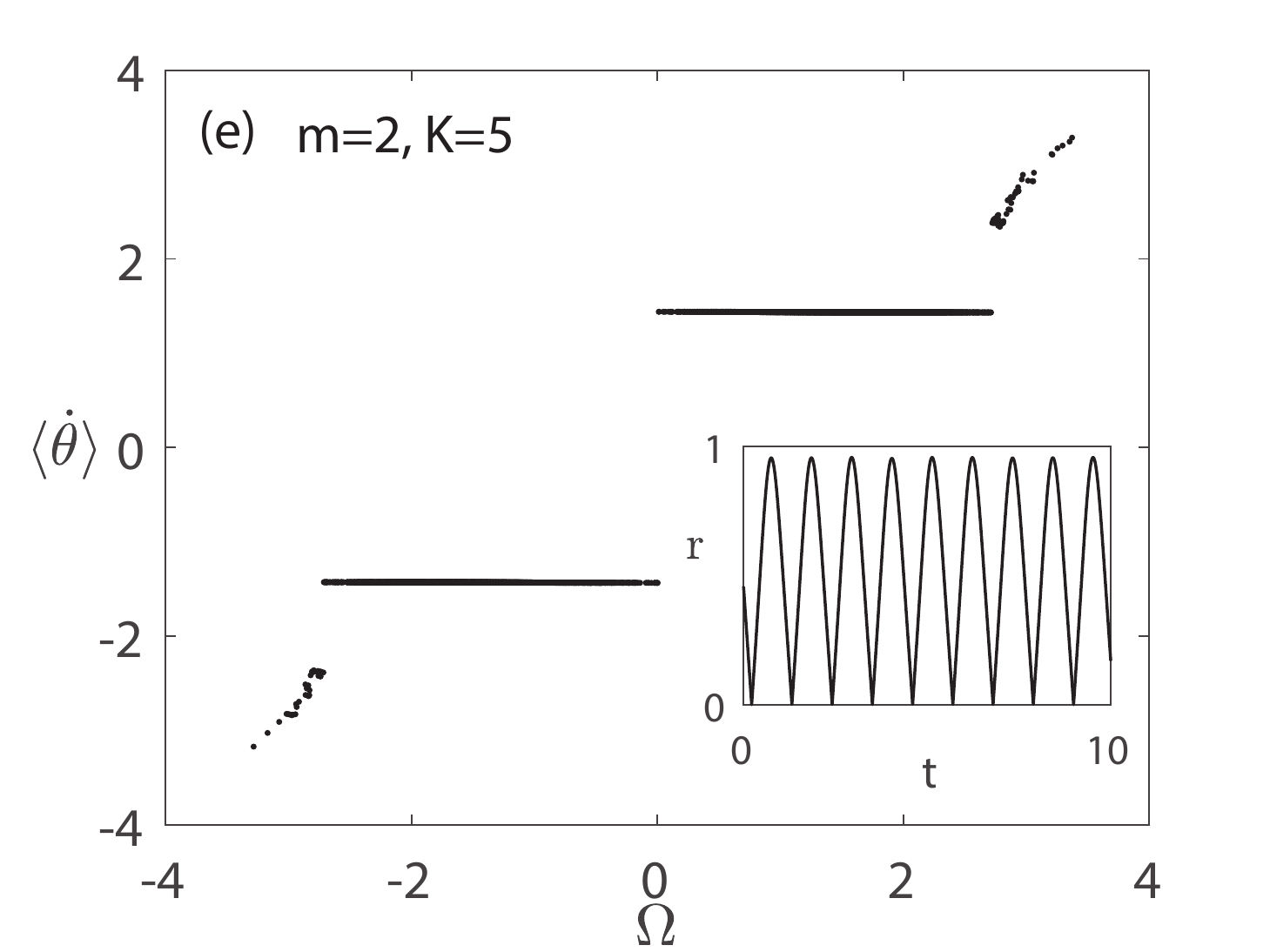}
\includegraphics[width=0.48\columnwidth]{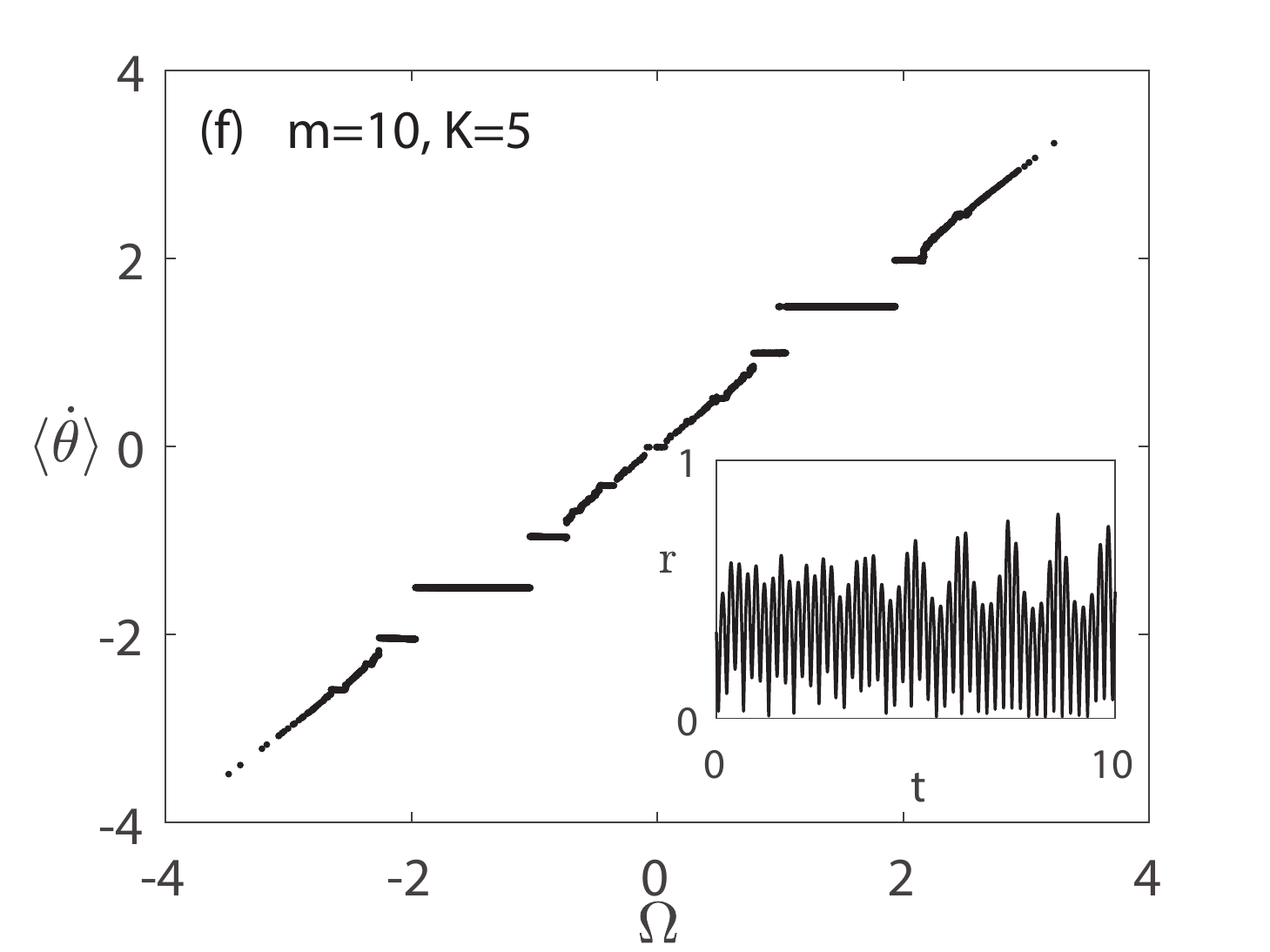}
\caption{Backward and forward processes for $N=1000$ oscillators with (left column) $m=2$ and (right column) $m=10$ for oscillators with bimodal natural frequency distribution $g_3(\Omega)$, Eq.~\eqref{eq_bimodal_narrow}.
In panels (e,f) two typical states are shown with the same coupling strength $K=5$.
Other details are the same as in Fig.~\ref{fig_unimodal}.} 
\label{fig_bimodal_narrow}
\end{figure}

\begin{figure}[tbp]
	\centering
	\includegraphics[width=0.48\columnwidth]{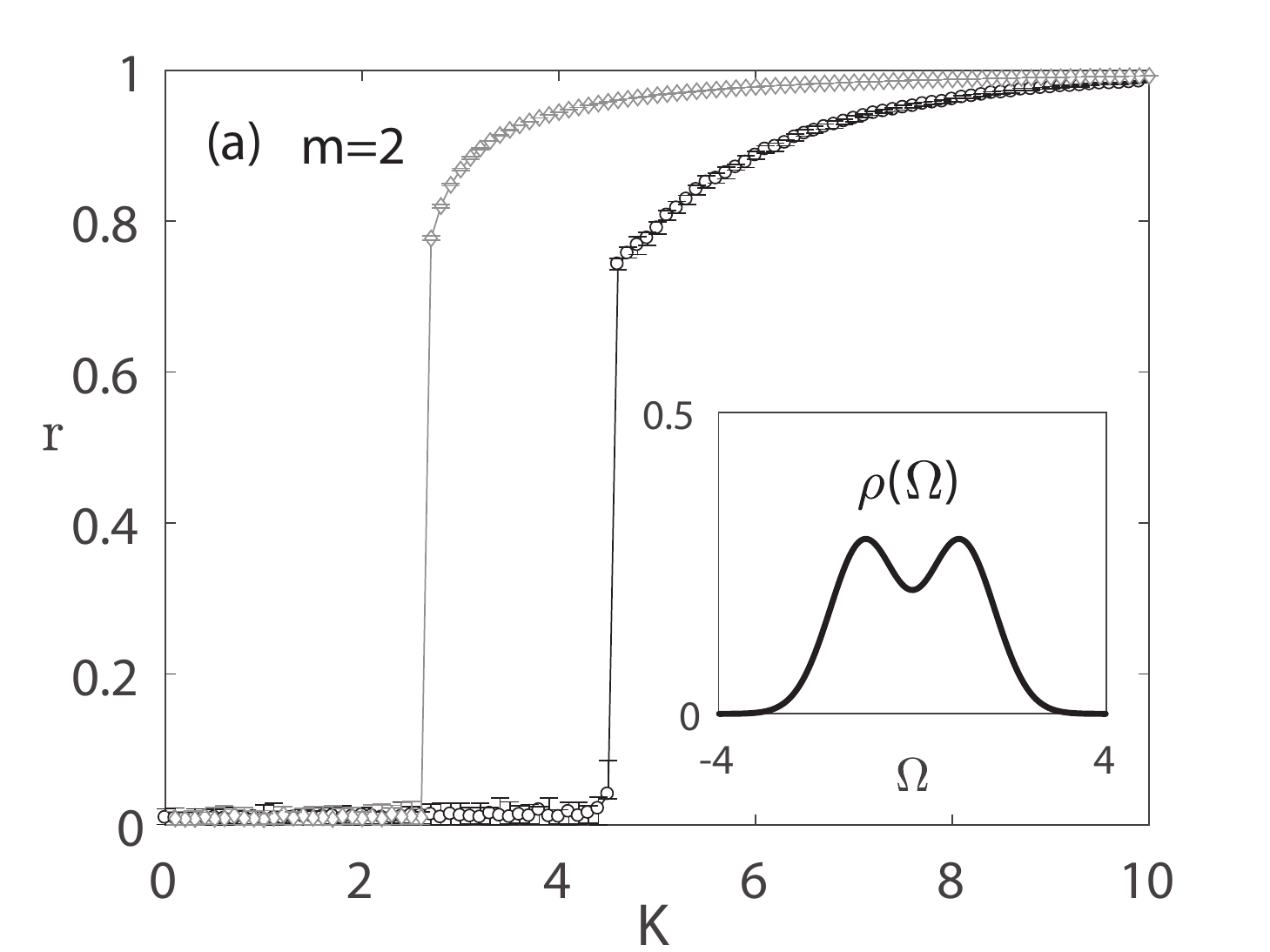}
	\includegraphics[width=0.48\columnwidth]{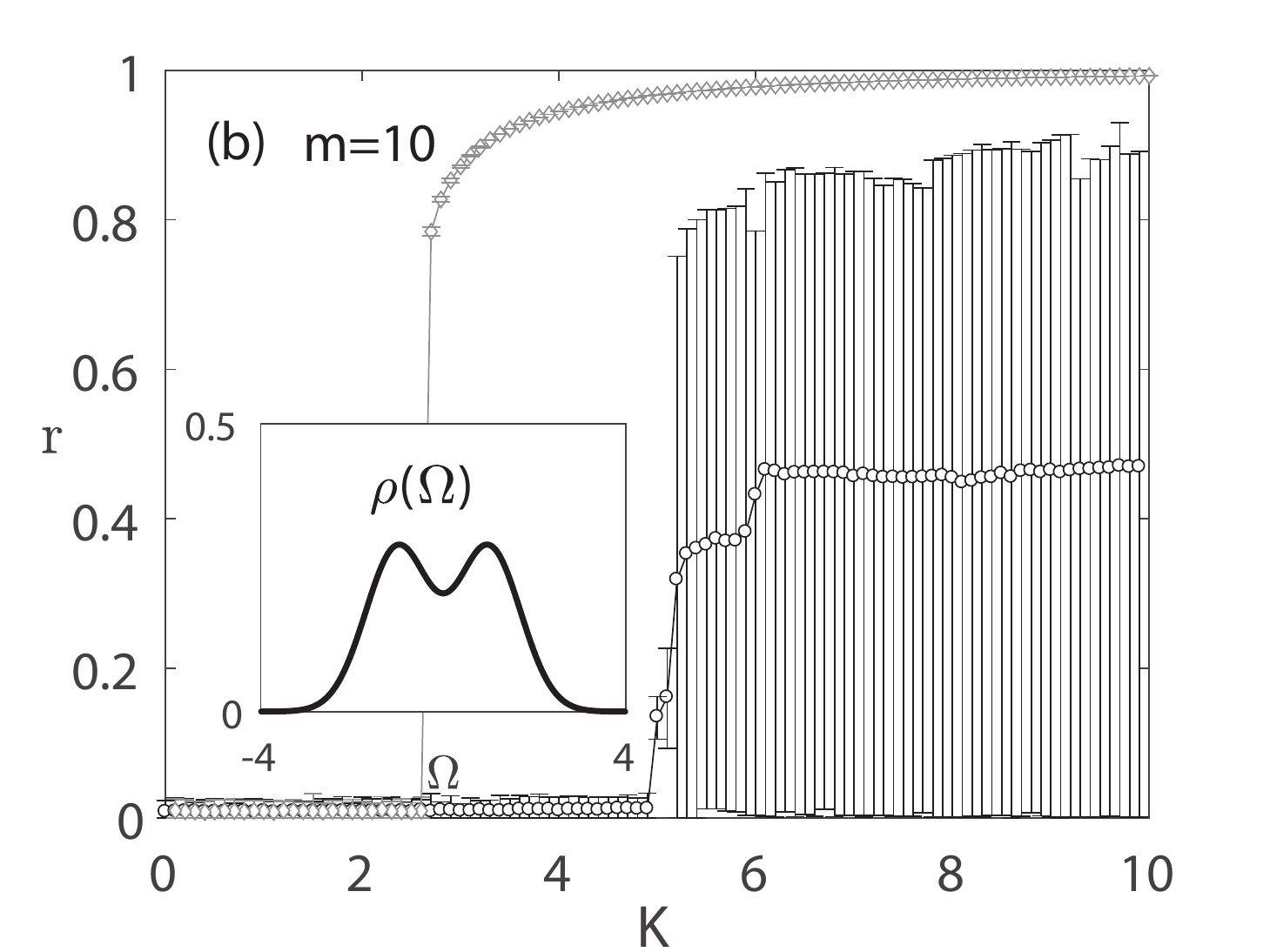}
	\includegraphics[width=0.48\columnwidth]{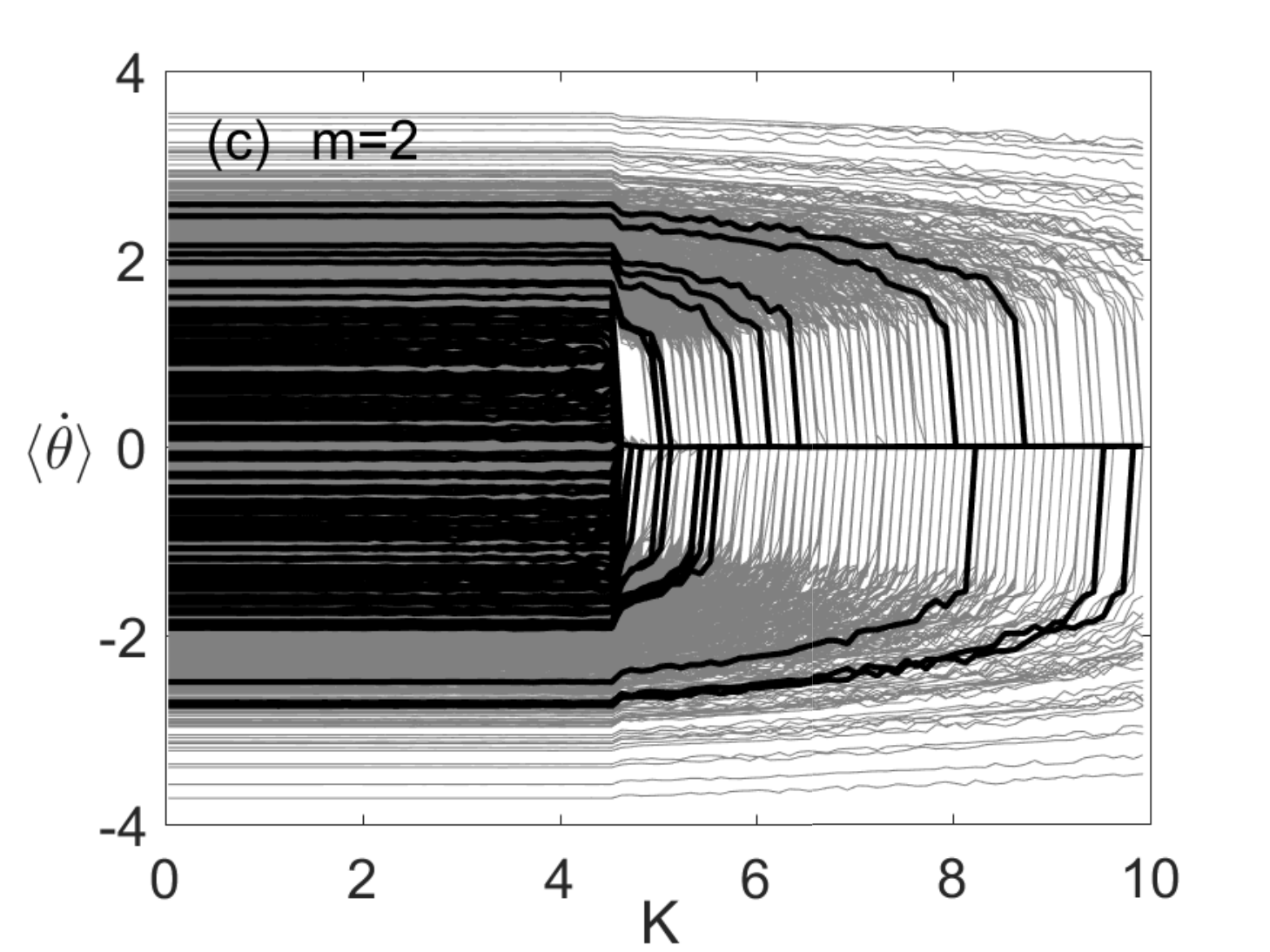}
	\includegraphics[width=0.48\columnwidth]{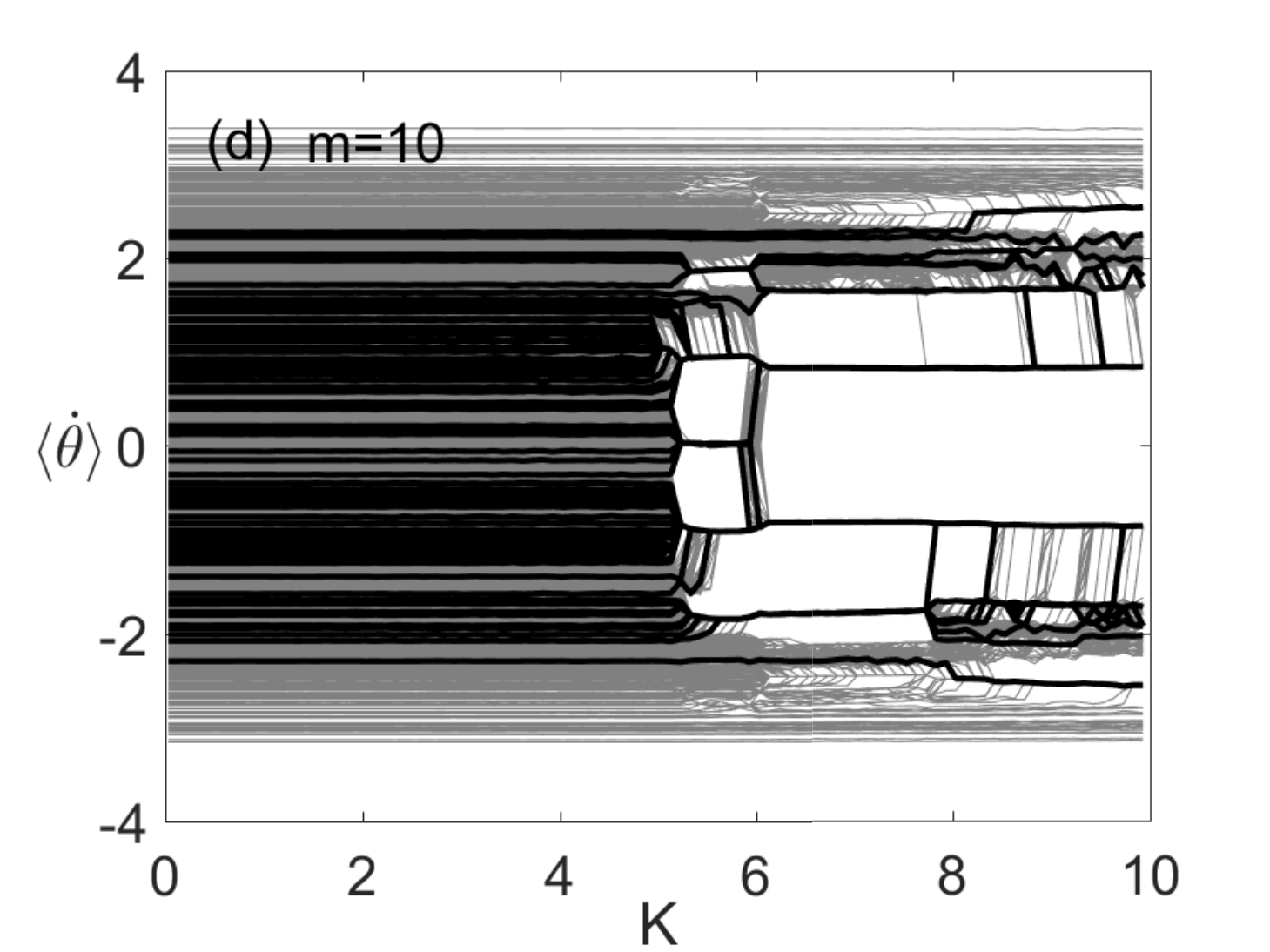}
	\includegraphics[width=0.48\columnwidth]{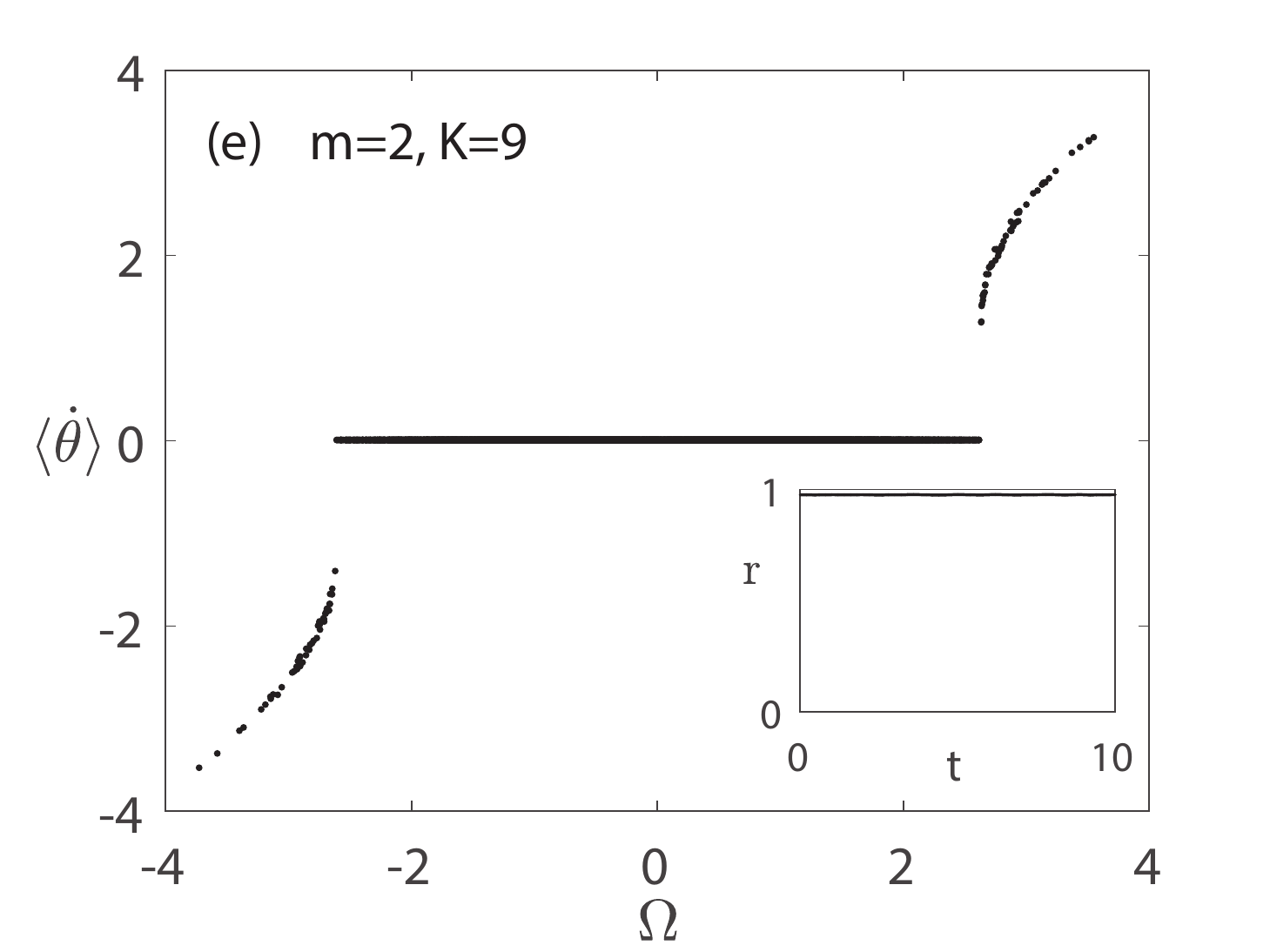}
	\includegraphics[width=0.48\columnwidth]{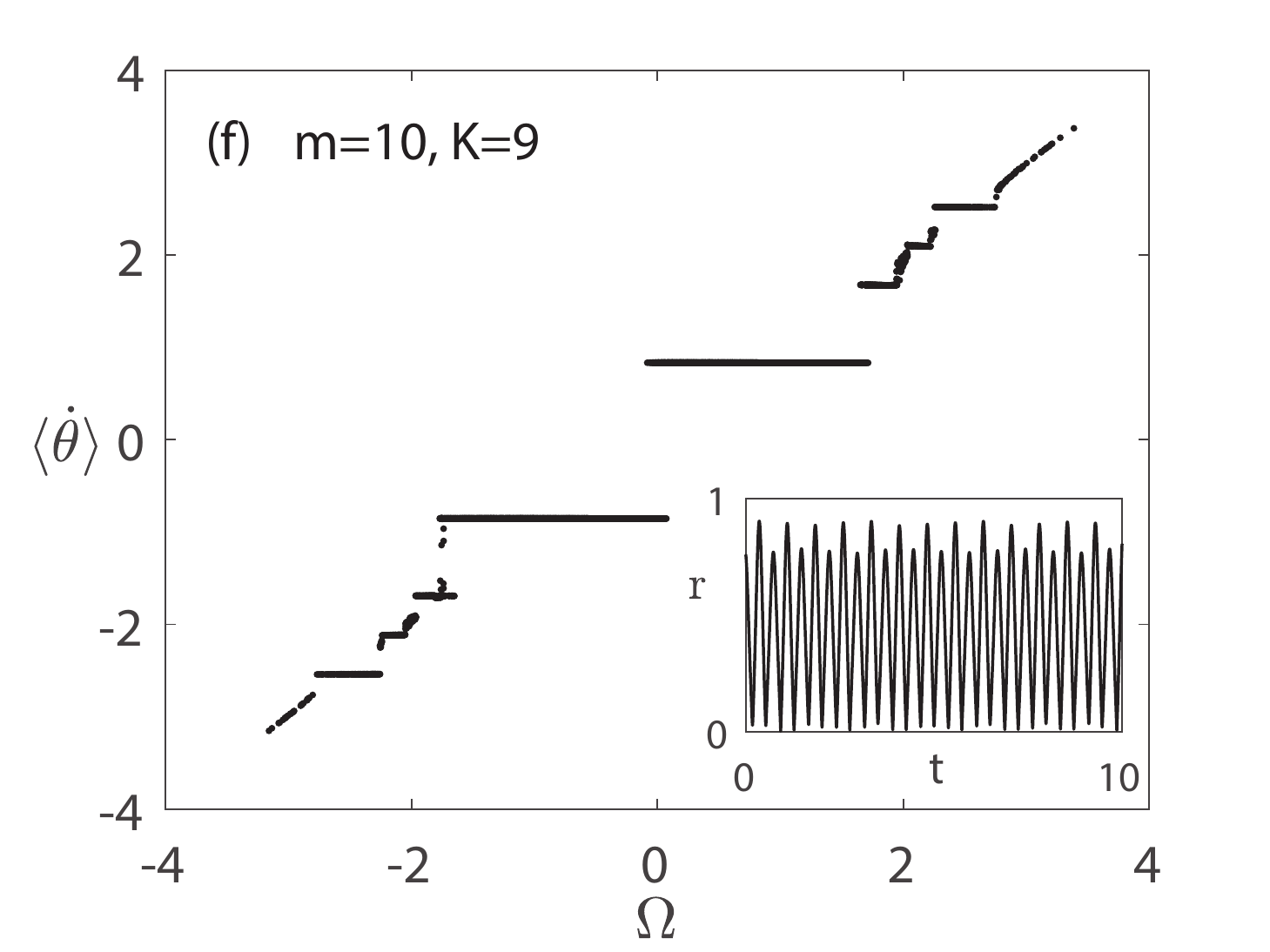}
	\caption{Backward and forward processes for $N=1000$ oscillators with (left column) $m=2$ and (right column) $m=10$ for oscillators with bimodal natural frequency distribution $g_2(\Omega)$, Eq.~\eqref{eq_bimodal_wide}.
		In panels (e,f) two typical states are shown with the same coupling strength $K=9$.
		Other details are the same as in Fig.~\ref{fig_unimodal}.} 
	\label{fig_bimodal_wide}
\end{figure}

In these calculations, we consider two effects:
the effect of inertia, $m$, and the effect of the natural frequencies distribution $g(\Omega)$.
In the first calculation oscillators with a Gaussian distribution
\begin{equation}\label{eq_unimodal}
  g_1(\Omega) = G(\Omega;0,1),
\end{equation}
where
\[
  G(\Omega;\mu,\sigma) = \frac{1}{\sqrt{2\pi \sigma^2}}
  \exp\left(-\frac{(\Omega-\mu)^2}{2\sigma^2}\right),
\]
are numerically explored with $m=2$ and $m=5$.
When $m=2$, the transition is discontinuous with a clear hysteresis, see Fig.~\ref{fig_unimodal}(a).
All the states are steady states, coinciding with the analytical calculation in \cite{Gao2018}.
For the larger $m=5$, depicted in Fig.~\ref{fig_unimodal}(b), the oscillators do not always reach a steady state and we have the appearance of oscillatory states comprising more than one synchronized clusters, as shown in Fig.~\ref{fig_unimodal}(d,f).
Note that the system still supports a steady state as predicted in \cite{Gao2018} but instead the oscillatory state is numerically observed.
We thus conjecture that the steady state becomes unstable leading to the appearance of the oscillatory state.
For both $m=2$ and $m=5$ there is a large synchronized cluster with average frequency $0$, see Fig.~\ref{fig_unimodal}(e,f). 
However, for $m=5$ we observe the appearance in Fig.~\ref{fig_unimodal}(f) of two additional synchronized clusters with non-zero average frequency at both sides of the main cluster.
These two clusters lead to the oscillations of the modulus of the order parameter, shown in the inset in Fig.~\ref{fig_unimodal}(f).
In previous studies, this state has been called \emph{secondary synchronization} \cite{Tanaka1997a} or \emph{oscillatory state} \cite{Olmi2014,Olmi2016}.
Fig.~\ref{fig_unimodal}(c,d) show how oscillators abruptly join the main synchronized cluster for $m=2$ when $K$ increases above the transition value, but they form additional synchronized clusters for $m=5$ even after $K$ goes above the transition value.

In the second calculation we consider oscillators with $m=2$ or $m=10$ and with a bimodal Gaussian distribution of natural frequencies
\begin{equation}
  g_3(\Omega) = \tfrac{1}{2} \big[ G(\Omega;1.5,0.5) + G(\Omega;-1.5,0.5) \big],
  \label{eq_bimodal_narrow} 
\end{equation}
where the two modes have a small overlap, see Fig.~\ref{fig_bimodal_narrow}. All 
the oscillators can be divided into two sub-groups with either positive or negative natural frequencies. For $m=2$ we observe the appearance of two sub-populations, whereas the $\avg{\dot\theta} = 0$ synchronized cluster is completely missing in contrast to the unimodal case where it was the most prominent one. The collective behavior of the system resembles two oscillators rotating in opposite directions where the order parameter oscillates along a constant or slowly varying direction. This is the \emph{standing wave} which is also observed for Kuramoto oscillators ($m=0$) with a bimodal natural frequency distribution \cite{Martens2009,Bonilla1998time}. For $m=10$, the oscillators are separated to two subgroups, with either negative or positive natural frequencies. For each of these subgroups, similar to the unimodal case, we observe the appearance in Fig.~\ref{fig_bimodal_narrow}(f) of two additional synchronized clusters at both sides of its main cluster. In addition, the $\avg{\dot\theta} = 0$ synchronized cluster is also observed in contrast to the case with $m=2$.  

Finally, in the third calculation we consider oscillators with $m=2$ or $m=10$ and with a bimodal Gaussian distribution of natural frequencies
\begin{equation}
g_2(\Omega) = \tfrac{1}{2} \big[ G(\Omega;1,0.7) + G(\Omega;-1,0.7) \big],
\label{eq_bimodal_wide}
\end{equation}
where the two modes strongly overlap, see Fig.~\ref{fig_bimodal_wide}.
For $m=2$, Fig.~\ref{fig_bimodal_wide}(a,c,e), the system behaves very similarly to the unimodal case, which can be expected from the study of Kuramoto oscillators with $m=0$.
For $m=10$, however, we observe states similar to a \emph{standing wave} Fig.~\ref{fig_bimodal_wide}(b,d), with several clusters besides this main structure. With the effect of inertias, the intrinsic two sub-groups structure is activated, forming a pair of synchronized clusters that rotate like two giant oscillators.

These numerical results indicate that the appearance of additional synchronized clusters is a general phenomenon of second-order oscillators.
In summary, the calculations indicate that increasing the inertia results in the appearance of additional synchronized clusters leading to oscillatory or standing wave states.
In addition, the phenomenon of additional synchronized clusters always appears in the lower branch of hysteresis loops (along the forward process), while the states in the upper branches (along the backward process) are not affected. 

\section{Time-periodic mean-field}
\label{sec/time-periodic-mean-field}

To understand the intrinsic synchronized clusters of the system, naturally the first step is to answer the question how synchronized clusters manifest under a given oscillatory mean-field.
For Kuramoto oscillators this question has only been recently addressed in \cite{Engelbrecht2012}.
For the second-order oscillators we consider here, the question becomes more complicated even though the idea behind our approach is similar to \cite{Engelbrecht2012}.

To analyze the oscillatory states, we first write the dynamics, Eq.~\eqref{eq_dynamics_original}, in mean-field form as
\begin{equation}\label{eq_dynamics_mean}
  m \ddot{\theta}_i + D\dot{\theta}_i
  = \Omega_i + K r(t) \sin(\Omega^r(t) t + \phi_0 - \theta_i),
\end{equation}
for $i=1,\dots,N$.
In Eq.~\eqref{eq_dynamics_mean} the oscillators interact with the mean-field through the order parameter.
In the particular case of oscillatory states we assume a time-dependent periodic mean-field modulus $r(t) = r_0 (1 + \varepsilon f(t))$ and constant $\Omega^r$.
Here $f(t)$ is a $T$-periodic function with zero average and $\varepsilon \ge 0$ measures the relative size of the time-dependent term.
With this assumption, the dynamics of oscillators can be written in mean-field form in a frame rotating as $\Omega^r t + \phi_0$ as
\begin{equation}\label{eq_dynamics}
  m \ddot{\theta} + D \dot{\theta} = \Omega - D\Omega^r
  - K r_0 (1+\varepsilon f(t)) \sin\theta.
\end{equation}

Further defining $\omega = \dot\theta$ we obtain on $M = \mathbb S \times \mathbb R$ the system of first-order differential equations
\begin{eqnarray}\label{Dynamics}
  \eqalign{\dot\theta &= \omega,} \\
  \eqalign{m \dot\omega &= - D \omega + (\Omega - D\Omega^r) - K r_0 (1+\varepsilon f(t)) \sin\theta.}
\end{eqnarray}
For a given initial state $(\theta(0), \omega(0))$, one can define the time-$T$ Poincar\'e map induced by Eq.~\eqref{Dynamics} as
\[
  F : M \to M : (\theta(0), \omega(0)) \mapsto (\theta(T), \omega(T)).
\]

\begin{figure}
\centering
  \includegraphics[width=0.48\columnwidth]{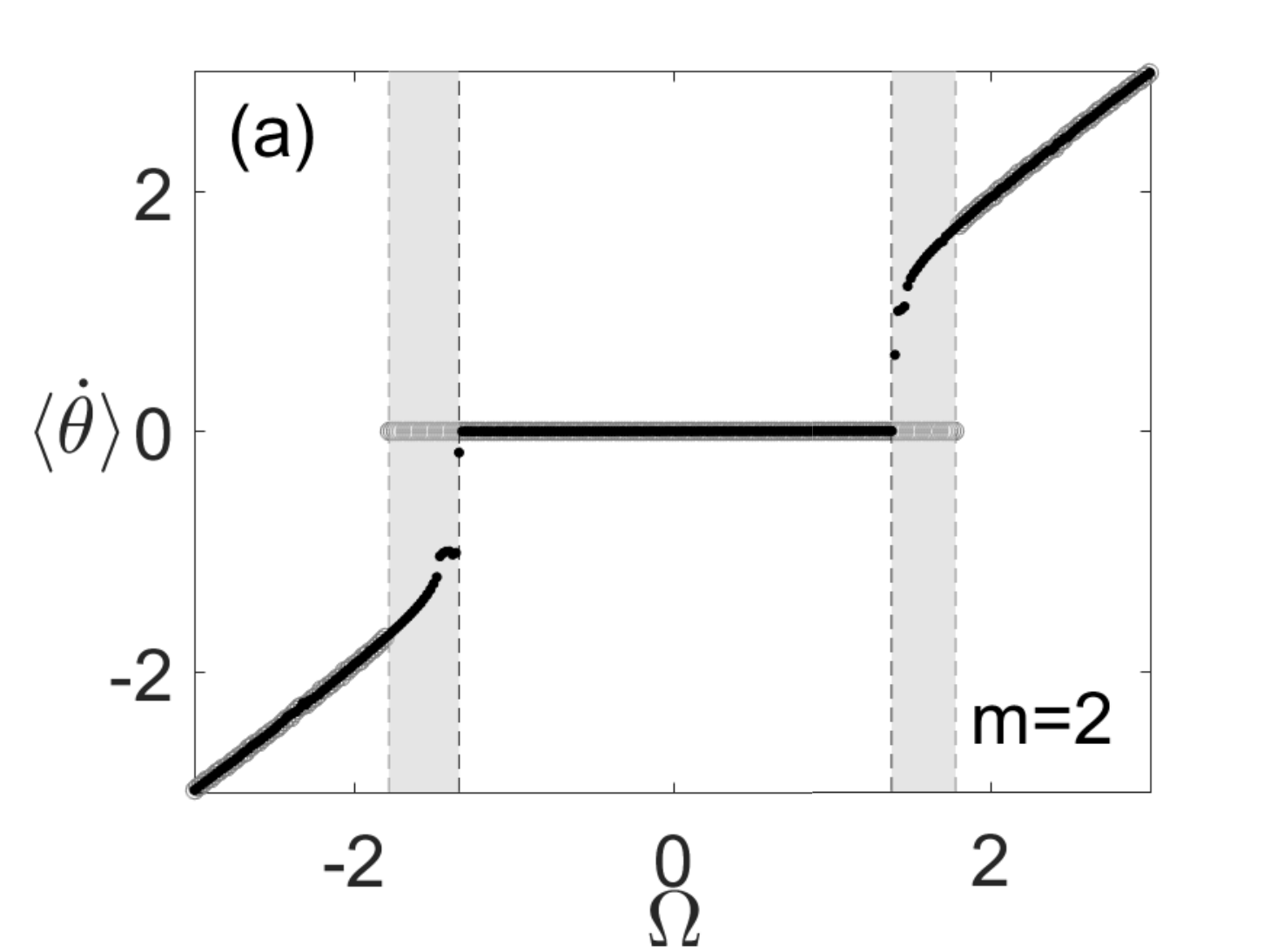}
  \includegraphics[width=0.48\columnwidth]{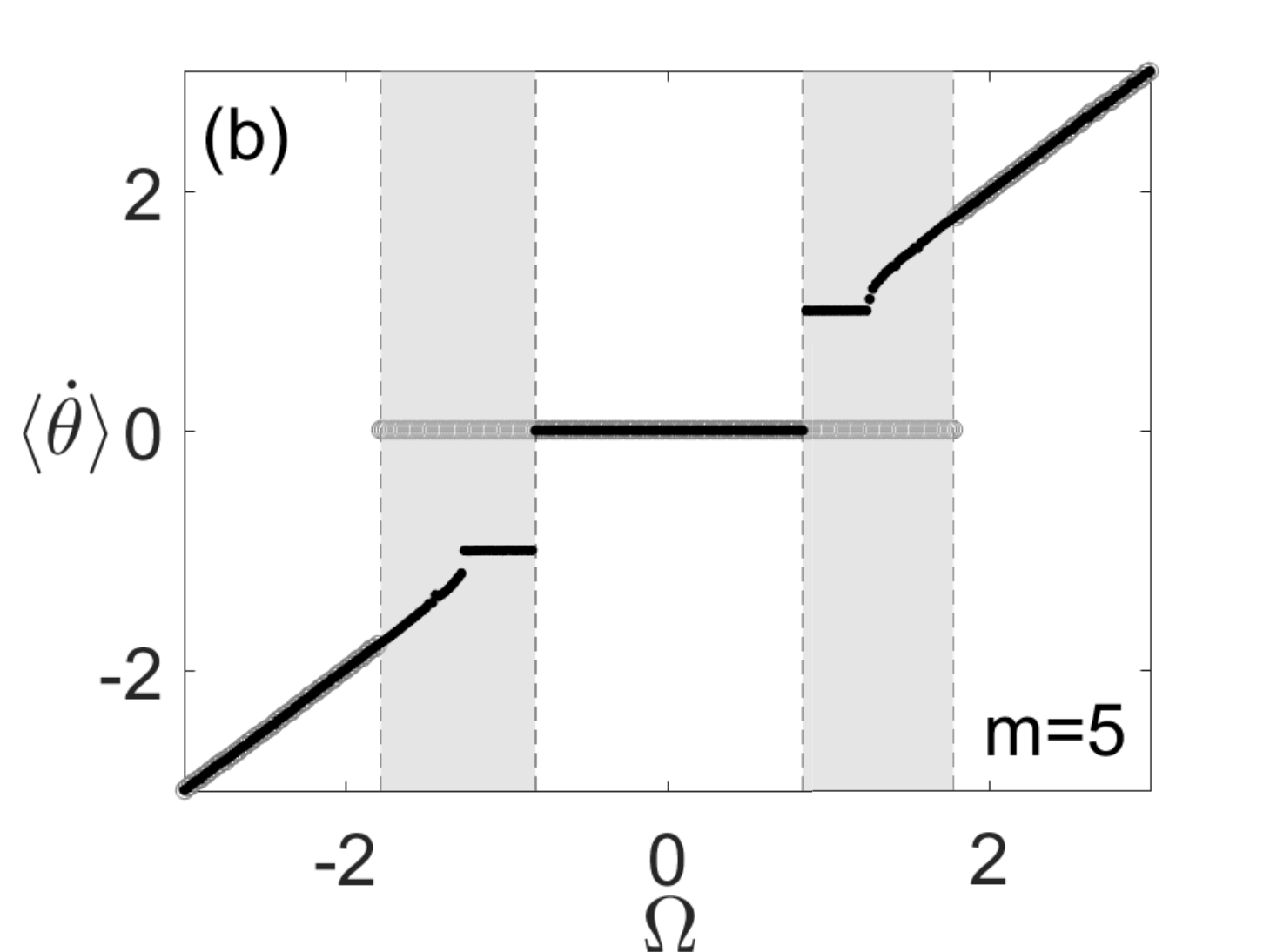}\\
  \includegraphics[width=0.48\columnwidth]{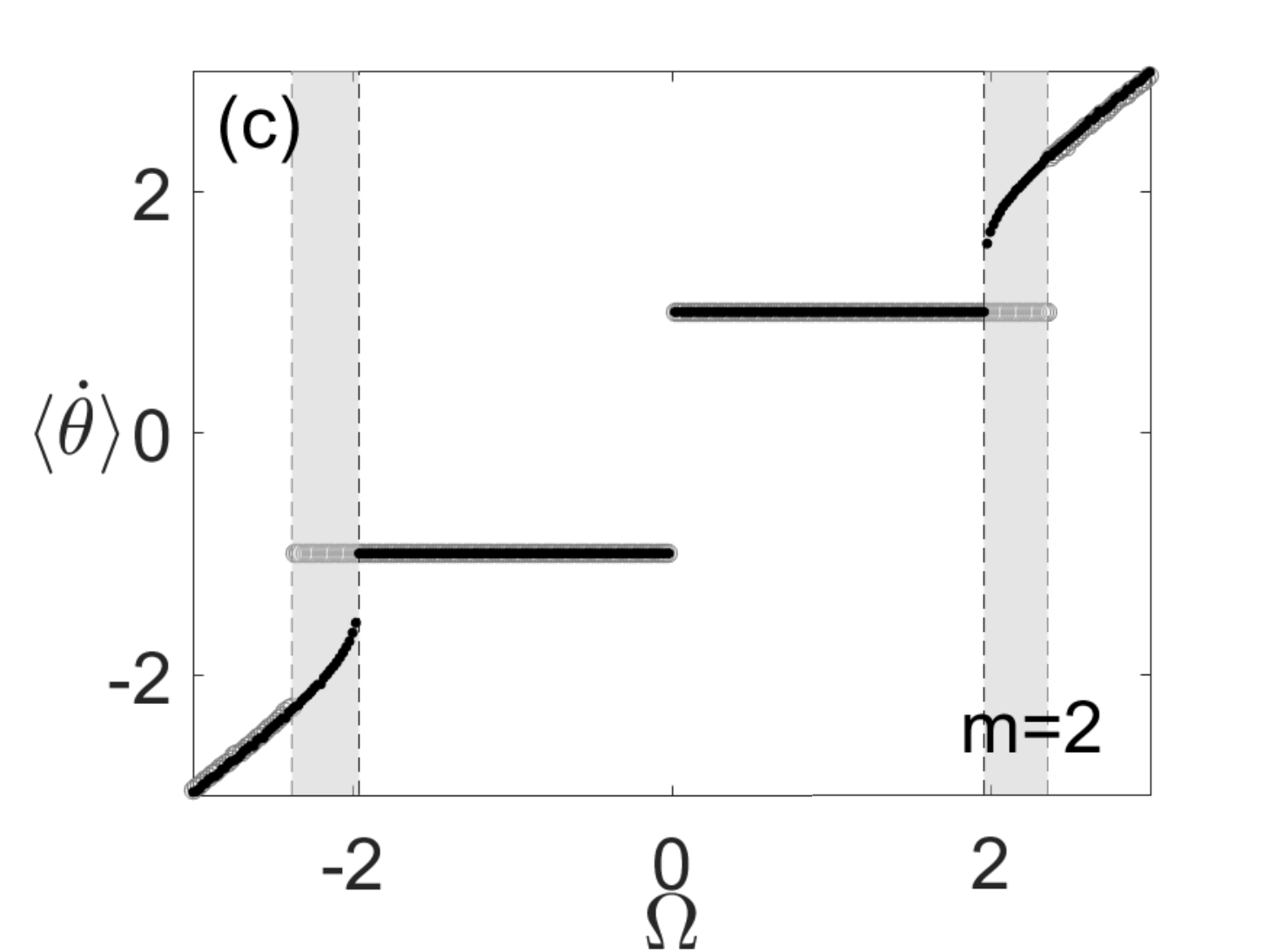}
  \includegraphics[width=0.48\columnwidth]{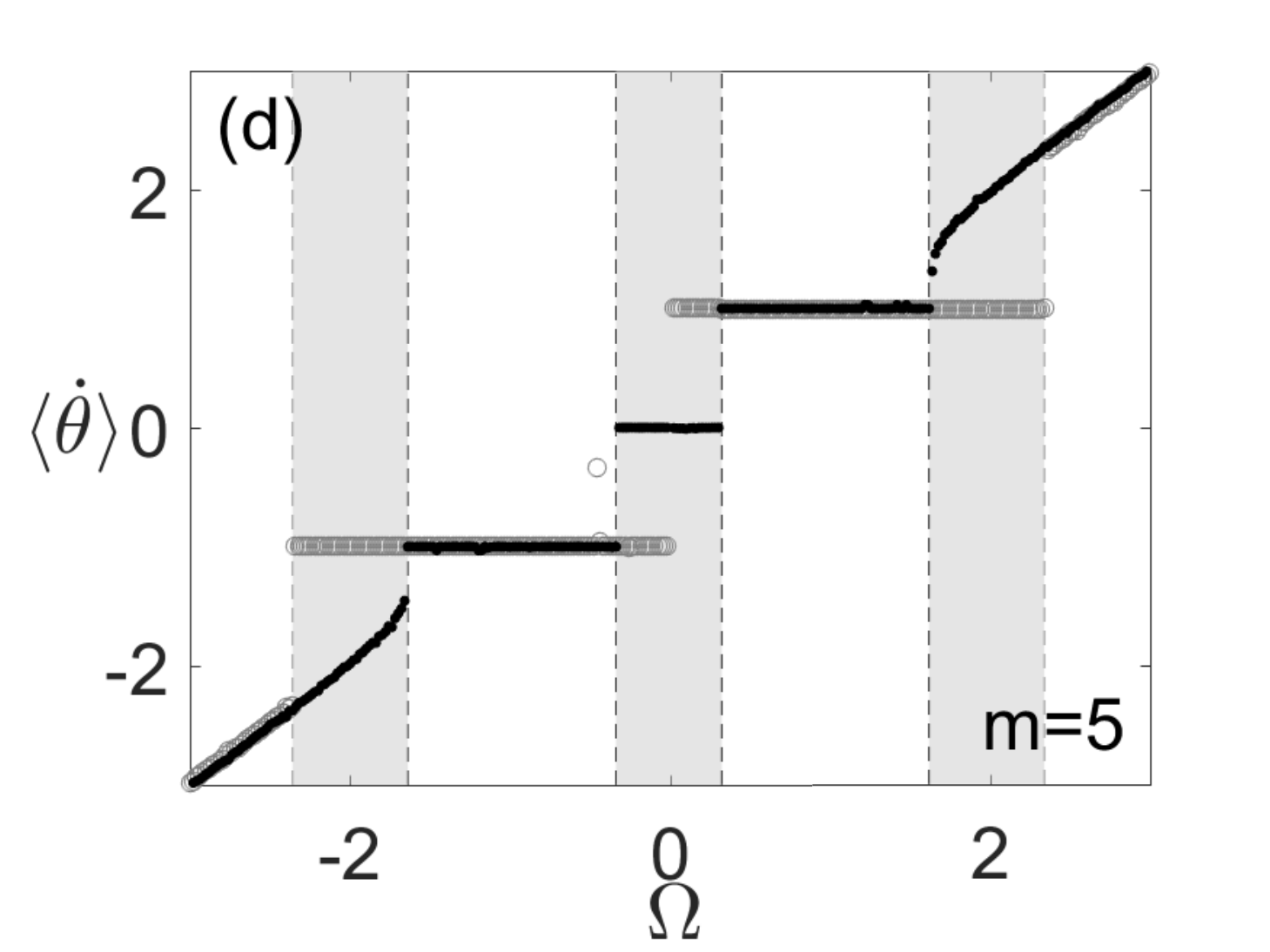}\\
  \caption{Mean-frequency $\langle\dot{\theta}\rangle$ with respect to natural frequency $\Omega$ with $m=1$ (a) and $m=5$ (b) where $r(t)=0.4+0.1\sin(t)$, and $m=1$ (c) and $m=5$ (d) where $r(t)=0.6\sin(t)$. The other parameters read $D=1, K=4.5$.
  The gray circles are typical states independent of different inertias (steady states for $r(t)=0.4+0.1\sin(t)$ and standing wave for $r(t)=0.6\sin(t)$), and the black dots are the states from the bi-stability of oscillators with inertia effect. The bistable regions are colored with gray.}
  \label{fig_bistable_region}
\end{figure}

The case $\varepsilon = 0$ corresponds to a steady state with $r(t) = r_0$.
In this case, Eq.~\eqref{eq_dynamics_original} has two possible stable states \cite{Levi1978,Strogatz2014,Guckenheimer2013,Gao2018}.
Introducing the parameters $a = D / (K r_0 m)^{1/2}$ and $b = (\Omega-D\Omega^r)/(Kr_0)$, it is known that for $b \ge b_L := 1$ the only stable state is a limit cycle $L$ where the motion has frequency $\Omega_L = \Omega/D - \Omega^r$.
For $b \le b_S(a)$ the only stable state is a fixed point $(\theta_0,0)$.
The bifurcation curve $b_S(a)$ is given by
\begin{equation}
b_S(a) \approxeq
\left\{\eqalign{
1.2732\,a - 0.3056\,a^3,  & \quad 0 \le a \le 1.193, \\
1, & \quad a \ge 1.193.
}\right.
\end{equation}
When $b_S(a) \le b \le b_L := 1$ the system is \emph{bistable}---the stable fixed point and stable limit cycle co-exist.
Several properties of second-order oscillators, such as the discontinuous phase transitions to synchronization and the corresponding hysteresis of steady states, are closely related to this bistability.

Consider the extended phase space $\widehat M = \mathbb S \times \mathbb R \times \mathbb S_T$ with coordinates $(\theta,\omega,t)$ where $t$ is viewed as a periodic variable in $\mathbb S_T := \mathbb R / T \mathbb Z$.
The stable fixed point $(\theta_0,0)$ becomes in $\widehat M$ a stable limit cycle $(\theta_0,0,t)$, or equivalently a fixed point $(\theta_0,0)$ of the Poincar\'e map $F$. 
Similarly, the stable limit cycle $L$ becomes in $\widehat M$ a stable limit torus $L \times \mathbb S_T$ carrying quasi-periodic motions with frequencies $\omega_1 = \Omega_L$ and $\omega_2 = 2\pi/T$ and manifests on the Poincar\'e section as an invariant curve $L_0$ of $F$ carrying a quasi-periodic circle map with rotation number
\begin{equation*}
  \rho_0 = \frac{\omega_1}{\omega_2} = \frac{\Omega_L T}{2\pi}.
\end{equation*}
Recall that the rotation number for an orbit of the Poincar\'e map $F$ with initial condition $(\theta_0,\omega_0)$ is
\begin{equation*}
  \mathrm{rot}(F) (\theta_0,\omega_0) \mathrel{:=}
  \lim_{n \to \infty} \frac{\theta_n - \theta_0}{2 \pi n}, 
\end{equation*}
where $(\theta_n,\omega_n) = F^n(\theta_0,\omega_0)$.

Both the stable limit cycle and the stable limit torus are compact normally hyperbolic invariant manifolds and thus by Fenichel's theory \cite{Fenichel1971,Hirsch1970} we expect that for sufficiently small $\varepsilon > 0$ these structures will persist.
In particular, the fixed point $(\theta_0,0)$ of $F$ persists as the fixed point $(\theta_\varepsilon, \omega_\varepsilon)$ while the invariant curve $L_0$ persists as the invariant curve $L_\varepsilon$.
Extending terminology from the case $\varepsilon = 0$ we will refer to oscillators converging to the fixed point as \emph{locked} and those converging to the invariant curve as \emph{running}, cf.~\cite{Gao2018}.

The restriction of the Poincar\'e map $F$ on the invariant curve $L_\varepsilon$ gives rise to a circle map with a rotation number $\rho_\varepsilon$ independent of the initial condition on the invariant curve \cite{Devaney2003introduction}. 
Consider now an ensemble of oscillators characterized by different $\Omega$ while the other parameters determining the dynamics, that is, $m$, $D$, $K$, $r_0$, $\Omega^r$, $\varepsilon$ and the $T$-periodic function $f(t)$ are the same.
Then the value of $\Omega$ determines whether the Poincar\'e map $F$ for Eq.~\eqref{Dynamics} has a fixed point, an invariant curve, or both.
If there is only a (stable) fixed point $(\theta_\varepsilon, \omega_\varepsilon)$ then all orbits will eventually converge to it and their rotation number will be $\mathrm{rot}(F)(\theta_0,\omega_0) = 0$.
Similarly, if there is only a stable invariant curve $L_\varepsilon$ then all orbits will have rotation number $\mathrm{rot}(F)(\theta_0,\omega_0) = \rho_\varepsilon$.
In the bistable case, where both a fixed point and an invariant curve co-exist, we will find some oscillators with rotation number $0$ and some oscillators with rotation number $\rho_\varepsilon$ depending on their initial condition, which determines if they converge to the fixed point or the invariant curve, respectively.
Therefore, a plot of $\mathrm{rot}(F)$ vs $\Omega$ for each oscillator will consist of three regions: one where all oscillators are running and have rotation number $\rho_\varepsilon$ (which however depends on $\Omega$ and exhibits the typical devil's staircase structure), one where all oscillators are locked with rotation number $0$, and the bistable region where some oscillators have rotation number $0$ and some have rotation number $\rho_\varepsilon$.

With a larger bistable region, we have more and larger plateaus in the graph of $\rho_\varepsilon$ vs $\Omega$ (corresponding to synchronized clusters) as shown in Fig.~\ref{fig_bistable_region}(a) and (b). With a larger bistable region and corresponding larger plateaus, the time-periodic mean-field can excite a larger oscillation of the order parameter of the oscillators. Taking the time-periodic mean-field as an oscillating perturbation around steady states, a sufficient large excited oscillation of the order parameter means the instability of such steady state and the formation of an oscillatory state. 

The size of the bistable region depends on the value of $a$ which in turn depends, for fixed $K r_0$, on the \emph{reduced mass} $\mu = m/D^2$.
In particular, for small $\mu$ we expect that there is no bistable region, while for sufficiently large $\mu$, there is a bistable region whose size increases with $\mu$.   
This observation explains why oscillatory states do not appear for small values of $\mu$.

In addition, in the backward process, all the oscillators in the bistable region have rotation number $0$ and they are not located on the plateaus where the rotation number is $\rho_\varepsilon \ne 0$. Therefore, in this case the appearance of bistable regions due to inertia does not contribute to the appearance of oscillatory states. This is the reason why backward processes are always similar with either large or small values of $\mu$ and do not support oscillatory states.

The inertia effect is not limited to the steady states. A special case of the previous analysis is when $r_0 = 0$ or very small and $\varepsilon$ is large. The first condition, $r_0 = 0$ or small, implies that the order parameter oscillates,  corresponding to the standing waves. The mean-field equation becomes
\begin{equation*}
  m \ddot \theta + D \dot \theta = (\Omega - D \Omega^r) - K \varepsilon f(t) \sin \theta.
\end{equation*}
For $\varepsilon = 0$, the system has a stable limit cycle $L$ with frequency $\Omega_L = \Omega/D - \Omega^r$.
As $\varepsilon$ starts increasing the limit cycle persists as an invariant curve $L_\varepsilon$. 
However, for larger values of $\varepsilon$ the dynamics on $L_\varepsilon$ will give rise to fixed points when $\mu$ is large enough, see Fig.~\ref{fig_bistable_region}(c) and (d). 
With the increase of inertia, locked oscillators can also appear in the purely oscillating mean-field (standing wave) together with the running oscillators that also exist for small inertias.
Note that this process is the opposite of the process that occurs in the case of oscillatory mean-field, see Fig.~\ref{fig_bistable_region}(a) and (b), where for small inertia we have only locked oscillators and with the increase of inertia we also observe the appearance of running oscillators.

These results, in particular the bi-stability of oscillators and devil's staircase structure of $\rho_\varepsilon$, explain why a periodic mean-field leads to the appearance of secondary synchronized clusters besides the cluster of locked oscillators. With the same time-periodic mean-field, the appearance of secondary synchronized clusters depends on the value of $\mu$ and the direction (backward and forward) of synchronization processes. Such analysis links the states of oscillators with the mean-field of the system and shows the inertia effect in these correlations. However, the mean-field of coupled oscillators is formed from the self-organization of all the oscillators. In the next section, to study the the self-organization processes, we will focus on a more detailed model.

\section{Self-organization processes}
\label{sec/three}

As a complement to the analysis based on the mean-field, this section is devoted to the dynamics of a few oscillators and their self-organization toward synchronization. Since all the oscillators are connected with each other, there are no topological differences and the only factor that affects their synchronization process is their natural frequencies' distribution $g(\Omega)$.
To mimic the effect of $g(\Omega)$, we introduce a weighted model of finitely many oscillators where the dynamics is given for $i = 1,\dots,N$ by
\begin{equation}\label{eq_weighted_model}
  m\ddot{\theta}_i + D\dot{\theta}_i
  = \Omega_i
  + \frac{K}{N}\sum_{j=1}^{N} a_j \sin(\theta_j-\theta_i).
\end{equation}  
Here $a_i$ is the weight, describing the fraction of oscillators with natural frequency $\Omega_i$, and is used to mimic the distribution of oscillators.

\begin{figure}
	\centering
	\includegraphics[width=0.24\textwidth]{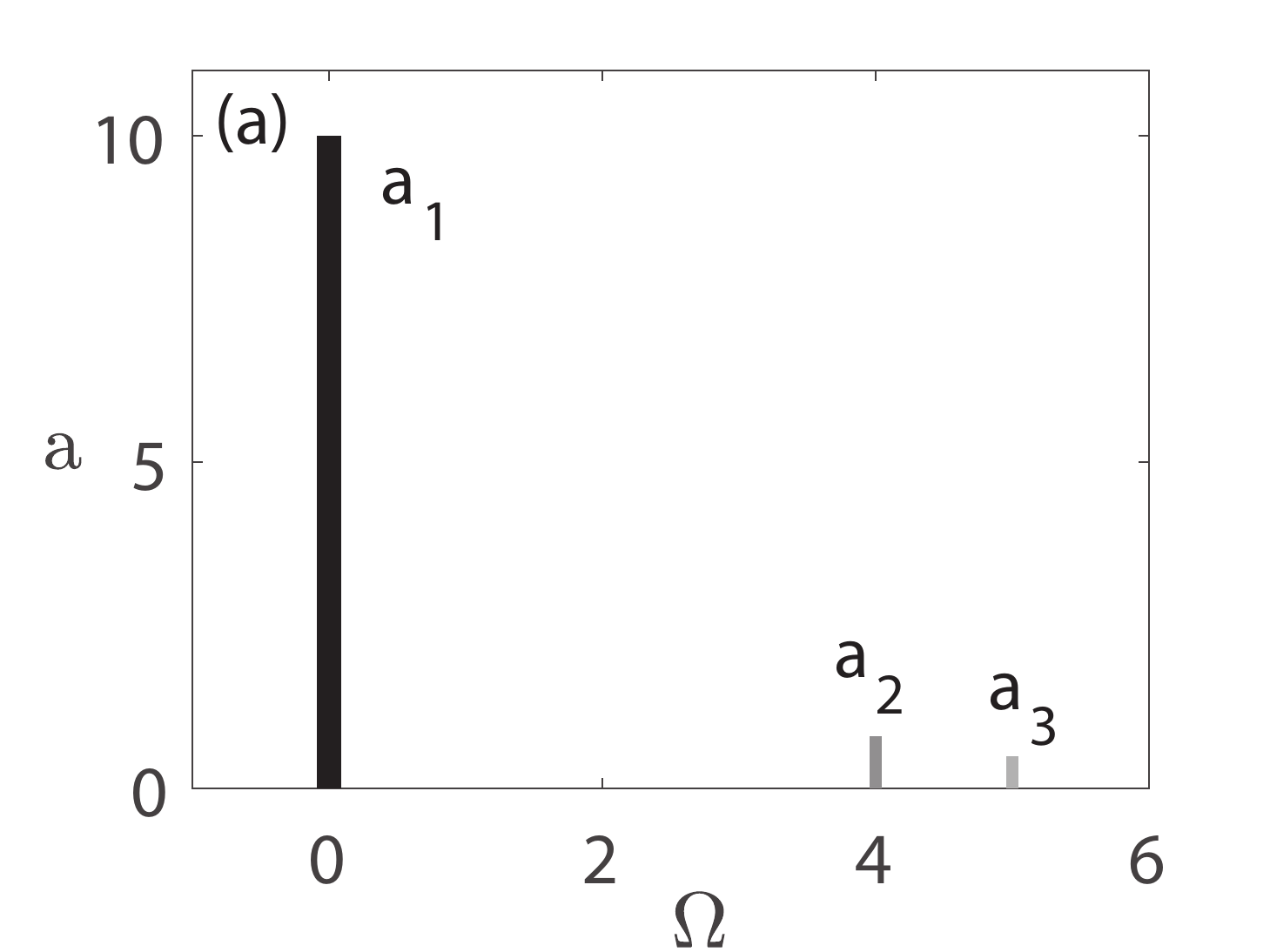}%
	\includegraphics[width=0.24\textwidth]{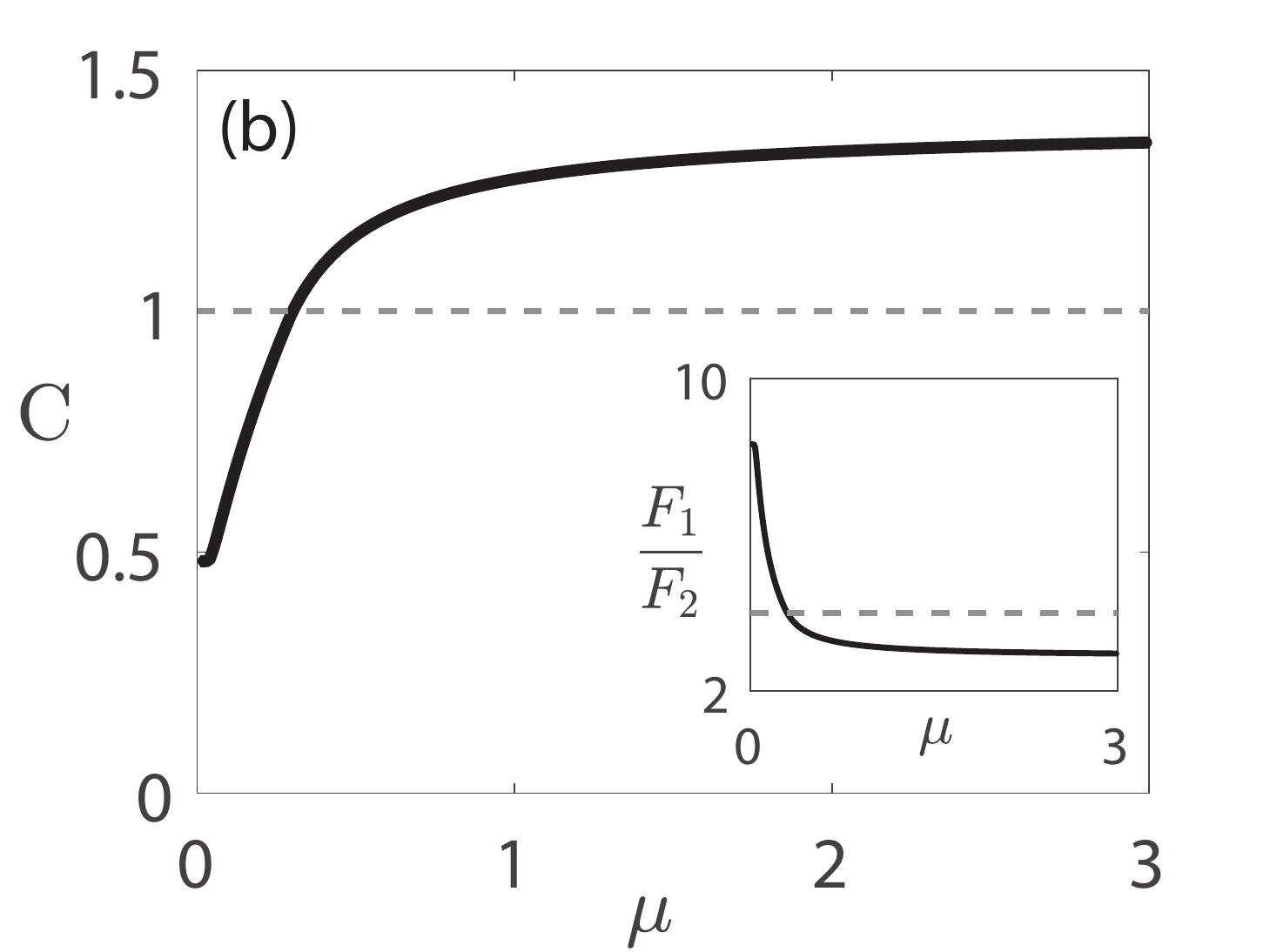}\\
	\includegraphics[width=0.24\textwidth]{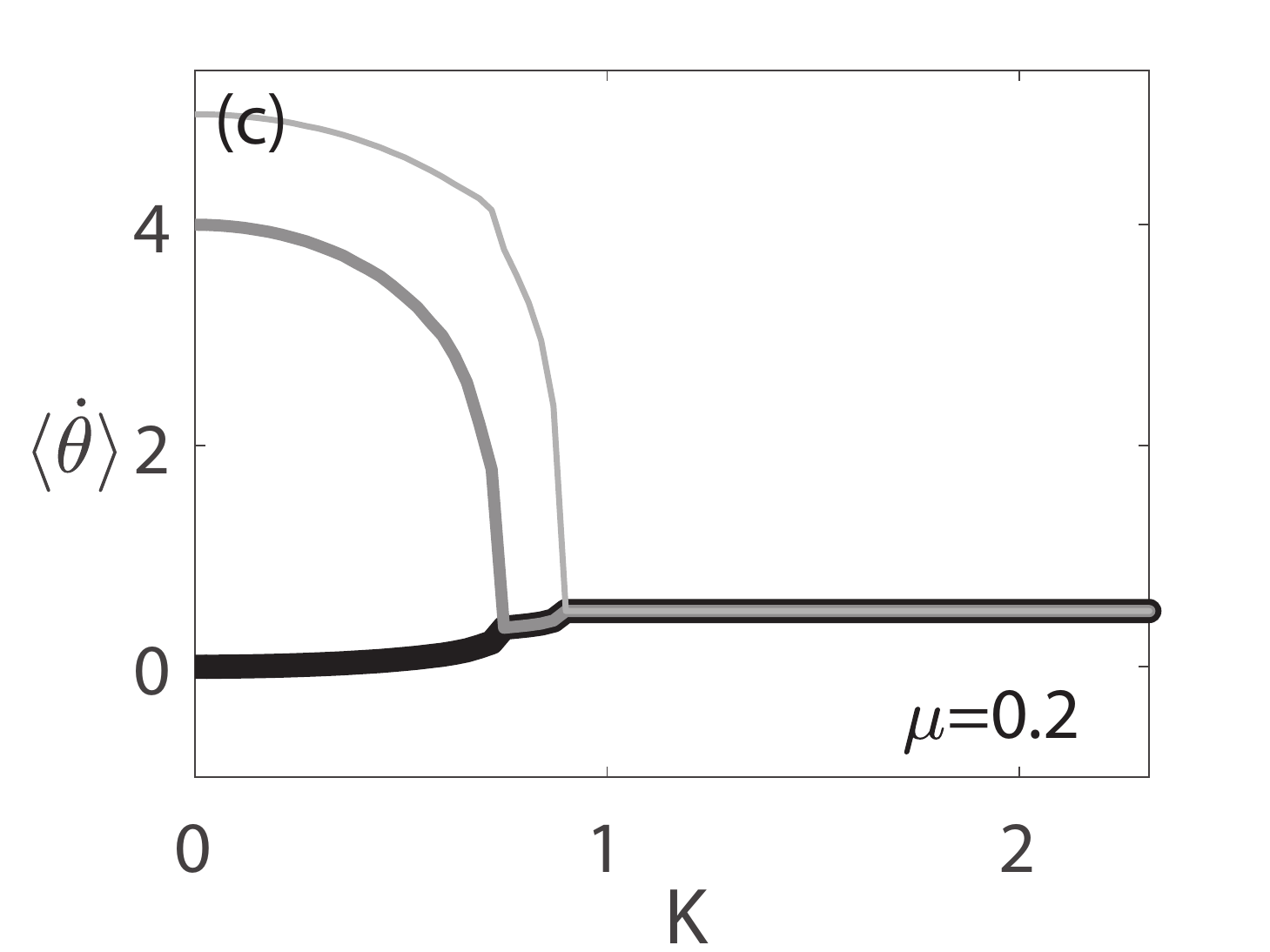}%
	\includegraphics[width=0.24\textwidth]{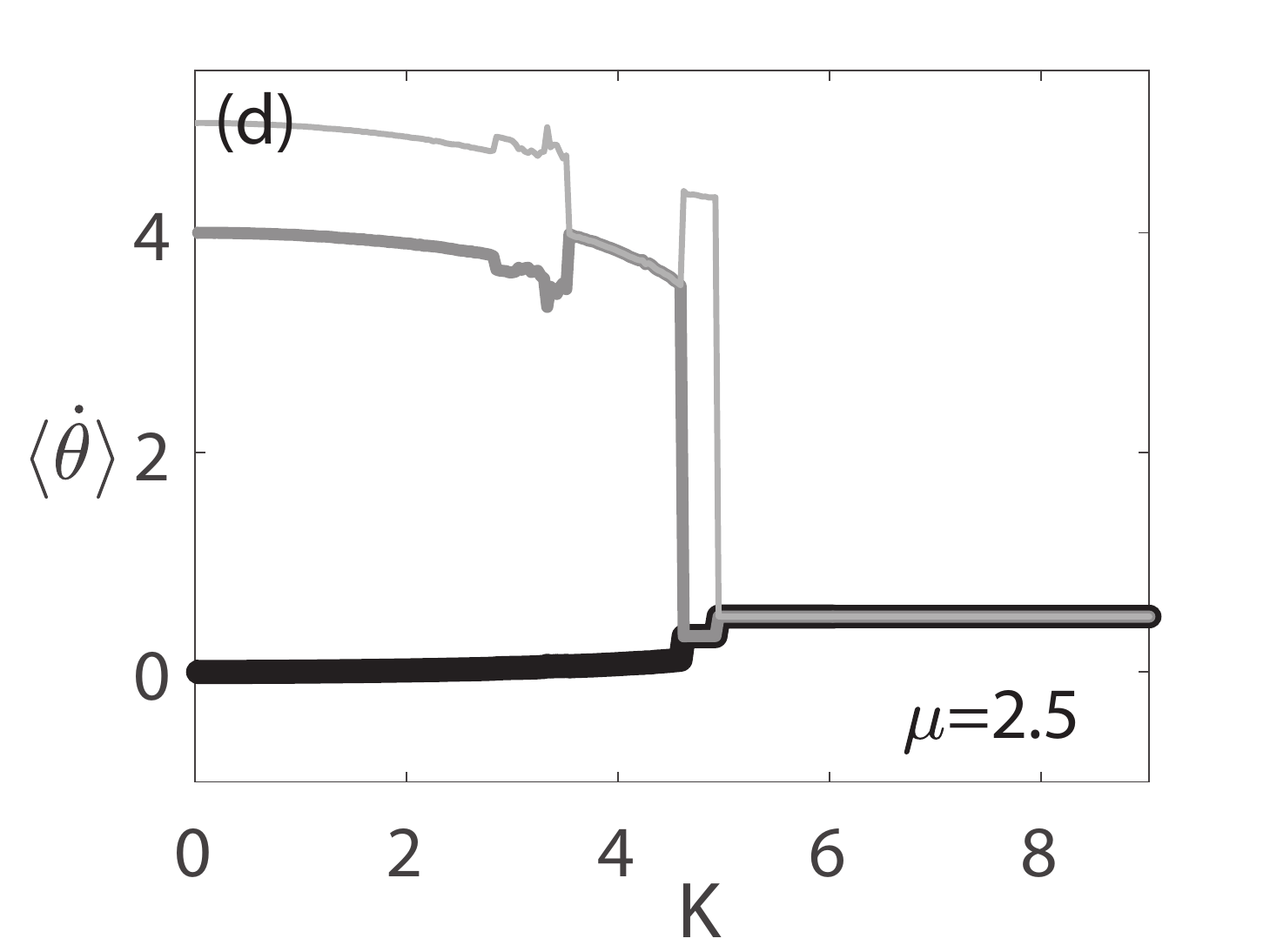}
	\caption{(a) Weights for oscillators with different natural frequency $\Omega$ as $a_1=10, a_2=0.8, a_3=0.5$. (b) The factor $C$ and $F_1/F_2\equiv F(a_{12})/F(a_{23})$ with different rescaled inertia $\mu$. (c) and (d) are the synchronization processes with increasing coupling $K$ with $m=0.2$ (b) and $m=2.5$ (d). The damping coefficient is $D=1$.}
	\label{fig_three_oscillator}
\end{figure}

To obtain more theoretical insights about the inertia effect, we begin by considering three coupled oscillators with $a_1\gg a_2>a_3$ and $\Omega_3>\Omega_2\gg\Omega_1$. Oscillator 1 is assigned the largest weight, describing the giant synchronized group, and oscillators 2 and 3 are assigned smaller weights but closer natural frequency, describing two small groups away from the giant group, see Fig.~\ref{fig_three_oscillator}(a). Introducing the phase differences $\varphi_1=\theta_1-\theta_2$, $\varphi_2=\theta_2-\theta_3$, we rewrite the dynamics as
\begin{eqnarray}
\eqalign{
m\ddot{\varphi}_1+D\dot{\varphi}_1 &= (\Omega_1-\Omega_2) - \frac{K}{2}
\Big[(a_2+a_1)\sin(\varphi_1) \cr & -a_3\sin(\varphi_2)+a_3\sin(\varphi_1+\varphi_2)\Big],}\\
\eqalign{
m\ddot{\varphi}_2+D\dot{\varphi}_2 &= (\Omega_2-\Omega_3) - \frac{K}{2} \Big[(a_3+a_2)\sin(\varphi_2)  \cr &-a_1\sin(\varphi_1)+a_1\sin(\varphi_1+\varphi_2)\Big].}
\label{eq_dynmaics_two}\end{eqnarray}

Even though the system Eq.~\eqref{eq_dynmaics_two} has various dynamical properties, we are only interested in the synchronization of each pair of oscillators, which means $\dot{\varphi}_1 \approxeq 0$ or $\dot{\varphi}_2 \approxeq 0$. The effect of the other oscillator is approximated as a periodic forcing by $\varphi_1\approxeq \omega_1t\equiv (\Omega_1-\Omega_2)/Dt$ when $\dot{\varphi}_2 \approxeq 0$ and $\varphi_1\approxeq \omega_2t\equiv (\Omega_2-\Omega_3)/Dt$ when $\dot{\varphi}_1 \approxeq 0$.

For the synchronization of the oscillators $\theta_1$ and $\theta_2$, we have 
\begin{equation*}
m\ddot{\varphi}_1+D\dot{\varphi}_1 =(\Omega_1-\Omega_2)-\frac{K}{2}\Big[(a_2+a_1)\sin(\varphi_1) 
-a_3\sin(\omega_2t)+a_3\sin(\varphi_1+\omega_2t)\Big],
\end{equation*}
From the fact that $a_1$ is the largest weight and $(a_2+a_1)\gg a_3$, we can ignore the periodic perturbations from $\varphi_2$ and obtain
\begin{equation}\label{eq_weight_first}
 m\ddot{\varphi}_1+D\dot{\varphi}_1=(\Omega_1-\Omega_2)-\frac{K}{2}(a_2+a_1)\sin(\varphi_1).
\end{equation}
As for the synchronization of the oscillators $\theta_2$ and $\theta_3$, we have
\begin{equation*}
m\ddot{\varphi}_2+D\dot{\varphi}_2
=(\Omega_2-\Omega_3)-\frac{K}{2}\Big[(a_3+a_2)\sin(\varphi_2) 
-a_1\sin(\omega_1t)+a_1\sin(\varphi_2+\omega_1t)\Big].
\end{equation*}
From the fact that $\omega_2\gg \omega_1$, we can average the fast periodic perturbation from $\varphi_1$ over time and obtain the dynamics of $\varphi_2$ as
\begin{equation}\label{eq_frequency_first}
m\ddot{\varphi}_2+D\dot{\varphi}_2=(\Omega_2-\Omega_3)-\frac{K}{2}(a_3+a_2)\sin(\varphi_2).
\end{equation}
Both the dynamics Eq.~\eqref{eq_weight_first} and Eq.~\eqref{eq_frequency_first} are the same as the dynamics of a single second-order oscillator Eq.~\eqref{eq_dynamics} with $\varepsilon=0$, studied in detail in \cite{Gao2018}. Hence the synchronization of each pair of oscillators, $\dot{\varphi_1}=0$ and $\dot{\varphi_2}=0$, can be obtained respectively as,
\begin{equation}\label{eq_synchronization_condition}
	\eqalign{
\frac{2(\Omega_1-\Omega_2)}{K(a_1+a_2) \, b\Big(\frac{\sqrt{2}}{\sqrt{K\mu(a_1+a_2)}}\Big)} &\equiv \frac{\Delta\Omega_{12}}{F(a_{12})}\leq 1, \cr
\frac{2(\Omega_2-\Omega_3)}{K(a_3+a_2) \, b\Big(\frac{\sqrt{2}}{\sqrt{K\mu(a_3+a_2)}}\Big)} &\equiv \frac{\Delta\Omega_{23}}{F(a_{23})}\leq 1,}
\end{equation}
where $F(a)=\frac{1}{2}K a\,b\big(\frac{\sqrt{2}}{\sqrt{K\mu a}}\big)$ is a function of the weight $a$. The boundary function $b(x)$ equals either $b_S(x)$ or $b_L(x)\equiv1$ in the forward and backward processes respectively. The frequency differences and sums of weights read $\Delta\Omega_{12}=\Omega_1-\Omega_2, a_{12}=a_1+a_2$ and $\Delta\Omega_{23}=\Omega_2-\Omega_3, a_{23}=a_2+a_3$. Since in the backward process the function $b_L(x)\equiv1$ is constant, the transition process is independent of $\mu$. However, in the forward process, with nonlinear boundary function $b_S(x)$, the synchronization processes depends on the value of $\mu$ crucially. 

To compare these two synchronization conditions, we define the factor
\begin{equation}\label{eq_condition}
  C = \frac{\Delta\Omega_{12}}{\Delta\Omega_{23}}\frac{F(a_{23})}{F(a_{12})}.
\end{equation}
If $C<1$ the dominant synchronization process is the growth of the giant group from oscillator 1, while if $C>1$ the additional synchronized cluster will form between oscillators 2 and oscillator 3.

The value of $C$ and hence the synchronization processes depends on the value of oscillators' rescaled inertia $\mu$, see Fig.~\ref{fig_three_oscillator}(b). When $\mu$ is small, we have that $F(a)=Ka/2$, and hence $F(a_{23})/F(a_{12})=a_{23}/a_{12}$. On the other hand, when $\mu$ is sufficient large the function $F$ can be approximated as $F(a)\approx \sqrt{8K/\mu\pi^2}\sqrt{a}$ and correspondingly $F(a_{23})/F(a_{12})\approx \sqrt{a_{23}/a_{12}}$. Even though the ratio $\Delta\Omega_{12}/\Delta\Omega_{23}$ is fixed, the ratio $F(a_{23})/F(a_{12})$ increases monotonically with the increase of $\mu$, from $a_{23}/a_{12}$ to $\sqrt{a_{23}/a_{12}}$, as shown in Fig.~\ref{fig_three_oscillator}(b). The effect of weights is weakened by the increase of inertias along the lower boundary $b_S$. Therefore, for larger inertias the oscillators are more likely to synchronize among ones with closer natural frequencies than with the giant group with larger weights. Consequently, we observe the appearance of additional synchronized clusters in the forward processes with sufficient large inertias as shown in Fig.~\ref{fig_three_oscillator}(d). As a generalization of these three-coupled oscillators, one can also consider three groups of oscillators as a limiting case of multimodal frequency distributions. This is beyond the scope of this paper and we refer to \cite{Acebron1998asymptotic,Acebron2001bifurcations}.

\begin{figure}
	\centering
	\includegraphics[width=0.24\textwidth]{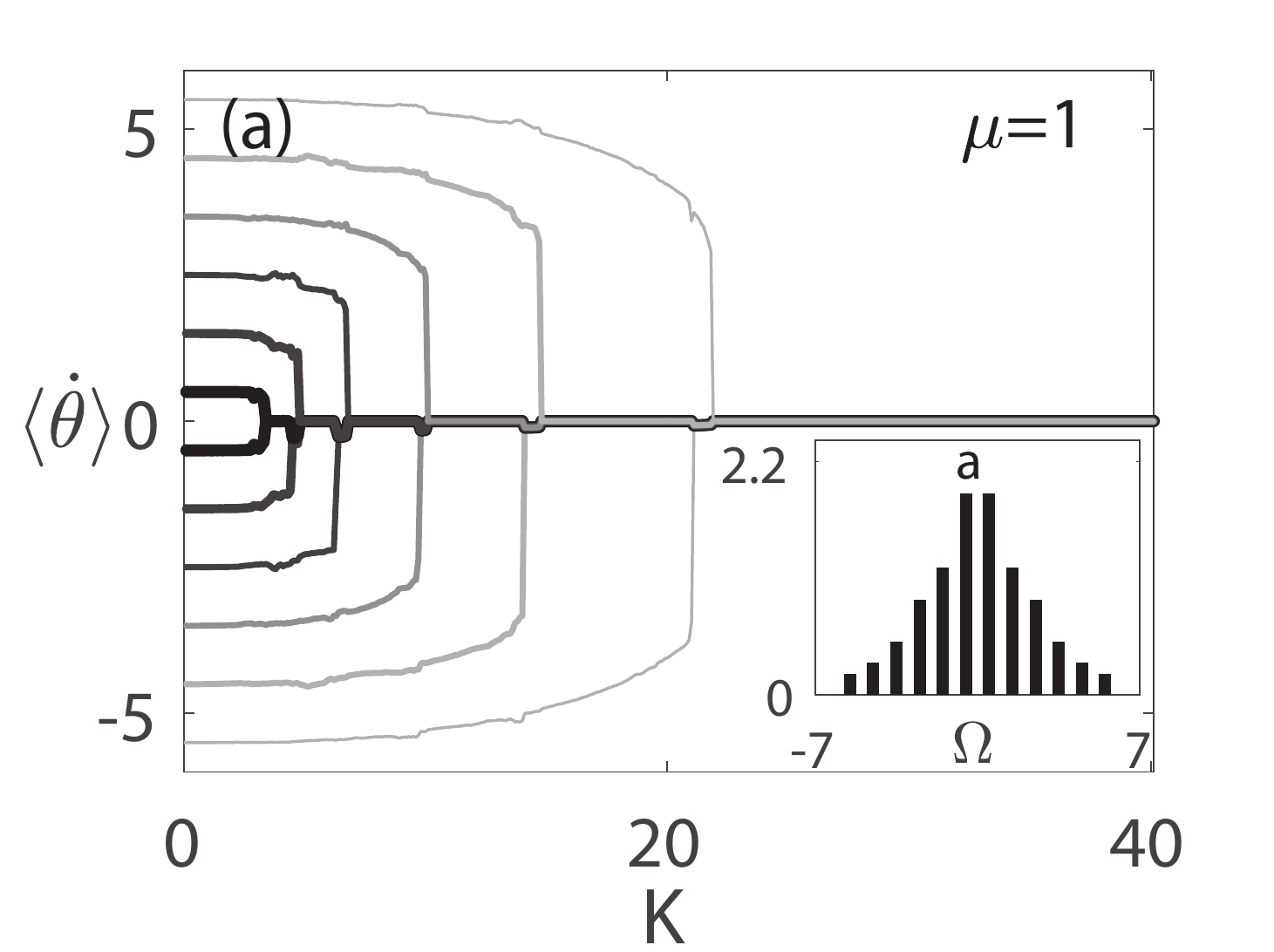}%
	\includegraphics[width=0.24\textwidth]{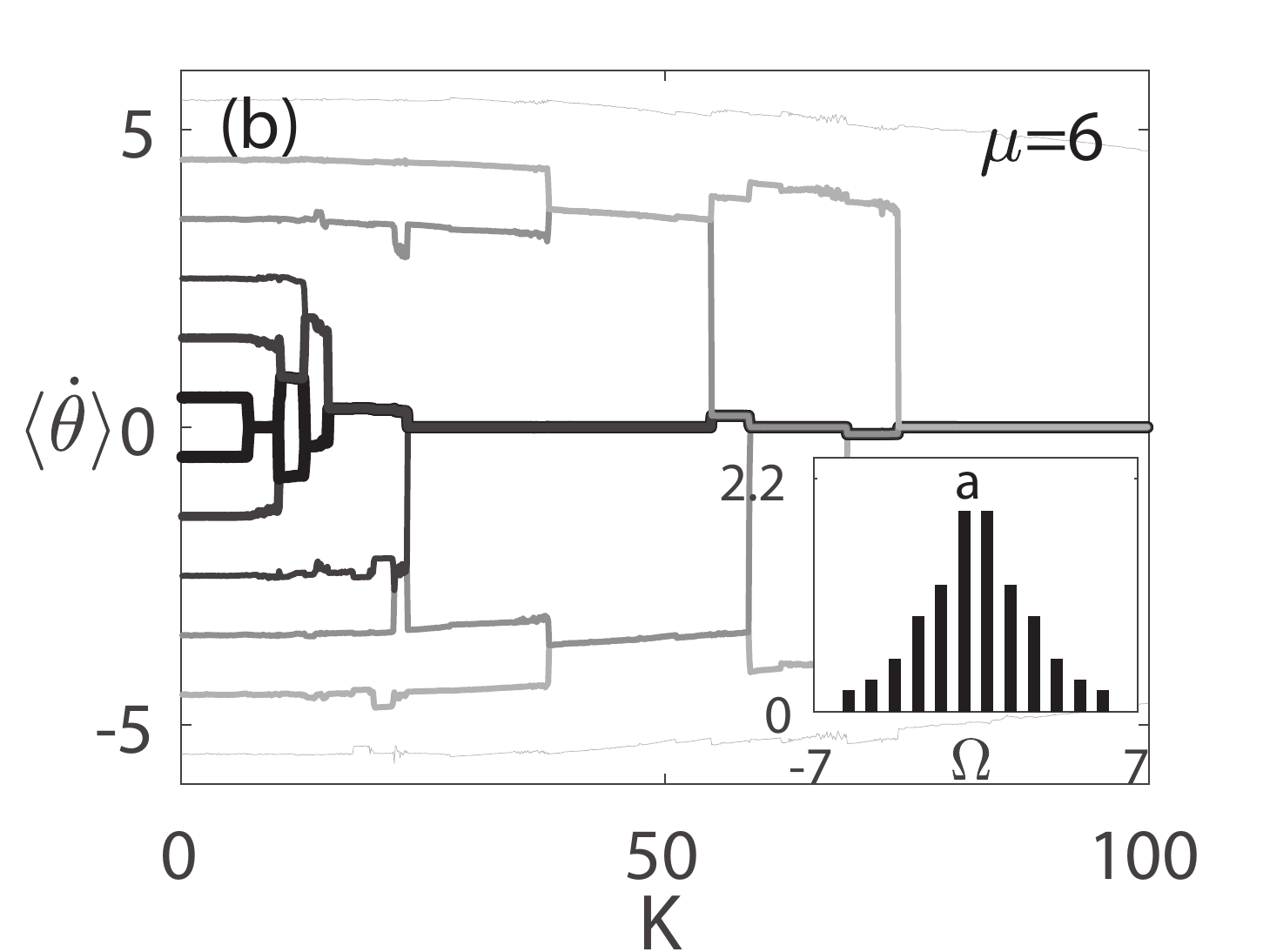}\\
	\includegraphics[width=0.24\textwidth]{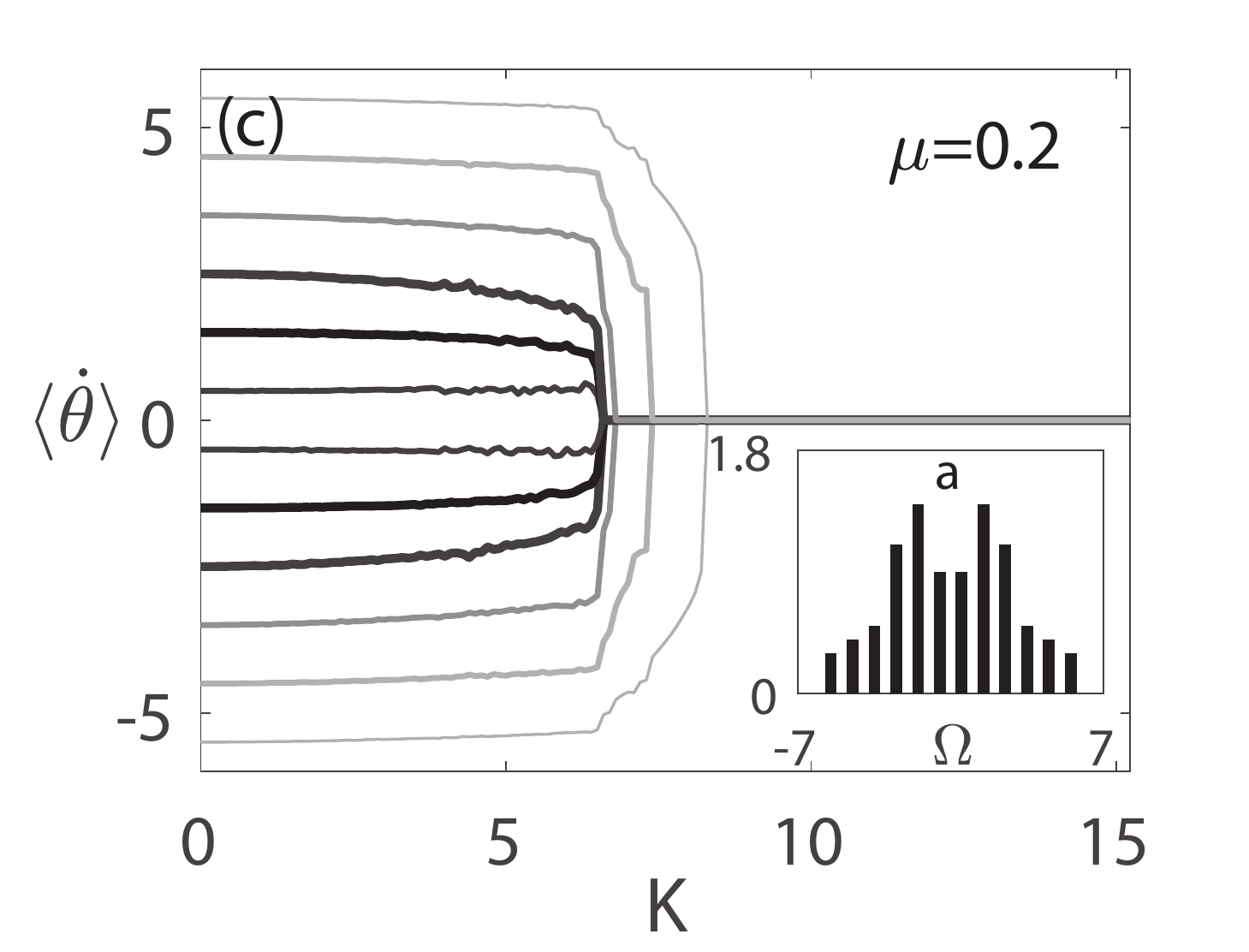}%
	\includegraphics[width=0.24\textwidth]{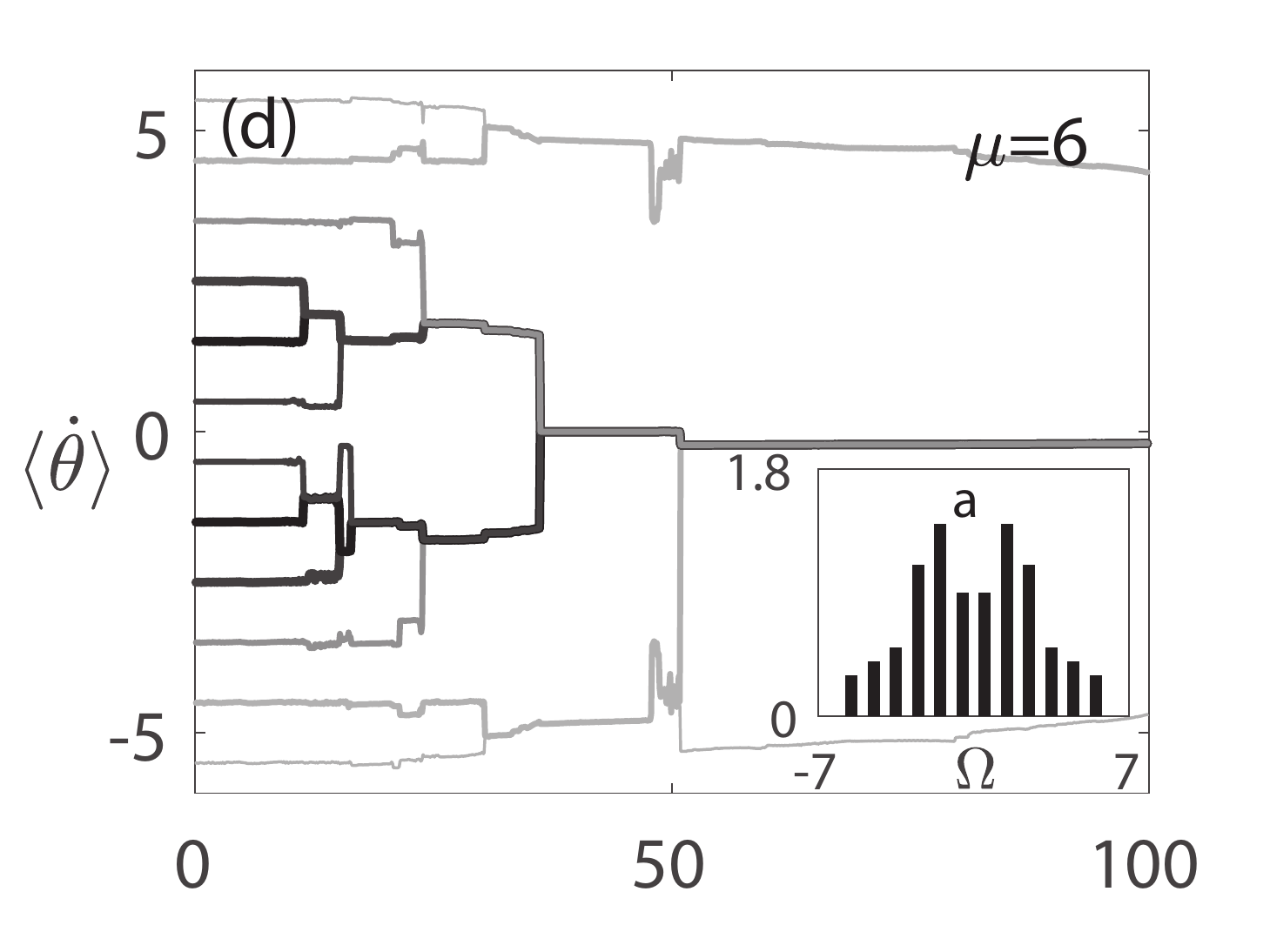}\\
	\includegraphics[width=0.24\textwidth]{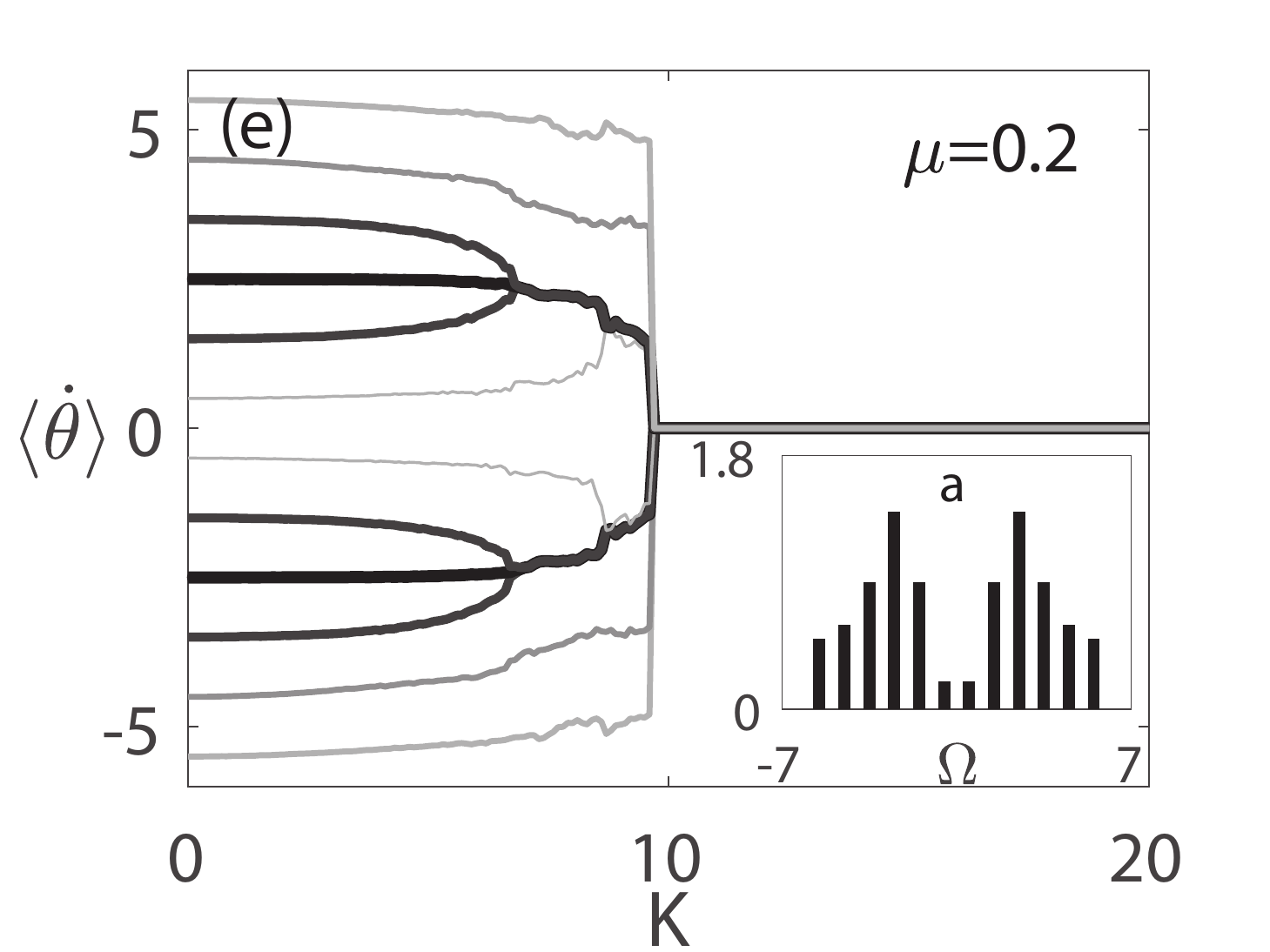}%
	\includegraphics[width=0.24\textwidth]{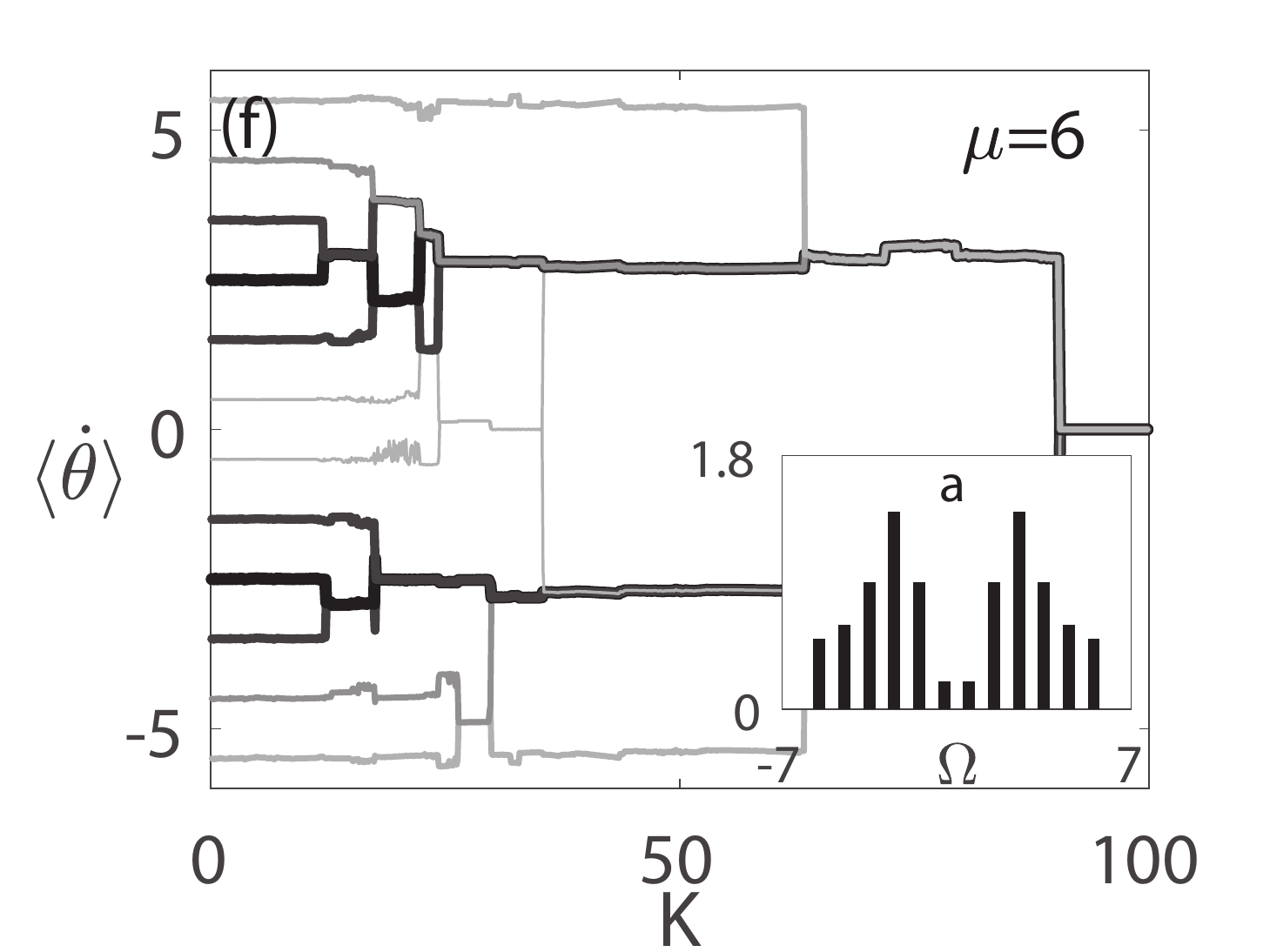}
	\caption{ Mean-frequency of $N=12$ oscillators with increasing coupling strength $K$ for unimodal distributed weights (a) ($\mu=1$) and (b) ($\mu=6$), large overlapped bimodal distributed weights (c) ($\mu=0.2$) and (d) ($\mu=6$), and small overlapped bimodal distributed weights (e) ($\mu=0.2$) and (f) ($\mu=6$). The natural frequencies are uniformly chosen as $\Omega_i=-5.5,-4.5,\dots,5.5$ with weights $a_i=0.2,0.3,0.5,0.9,1.2,1.9,1.9,1.2,0.9,0.5,0.3,0.2$ in (a) and (b), $a_i=0.3,0.4,0.5,1.1,1.4,0.9,0.9,1.4,1.1,0.5,0.4,0.3$ in (c) and (d), $a_i=0.5,0.6,0.9,1.4,0.9,0.2,0.2,0.9,1.4,0.9,0.6,0.5$ in (e) and (f).}
	\label{fig_12_oscillator}
\end{figure}

Following the analysis above for three oscillators, we further consider a larger system, where the synchronization process is a more complicated self-organization progress. Considering coupled $2N$ oscillators, with the dynamics Eq.~\eqref{eq_weighted_model} and uniformly spaced frequency $\Omega_i$ and different weights $a_i$, we could find different synchronization processes and states, see Fig.~\ref{fig_12_oscillator}.

Firstly, for a system with a symmetric and unimodal natural frequency distribution, the weights of the central pair of oscillators $\theta_N$ and $\theta_{N+1}$ have the largest weights. 
From the fact that all nearest pairs of oscillators have the same natural frequencies difference $\Delta\Omega$, the central pair with the largest weights will synchronize first in the process of increasing coupling strength $K$ from Eq.~\eqref{eq_condition}. Once they are synchronized, they form a synchronized cluster or equivalently an effective oscillator $\theta_0$ with the frequency $\Omega_0\equiv(\Omega_N+\Omega_{N+1})/2$ with weights $a_0\equiv(a_{N}+a_{N+1})$.

With the further increase of coupling strength, for all the oscillators with $\theta_i$ with $i>N$ the next synchronization phenomenon will happen either between oscillators $\theta_{N+2}$ with $\theta_{N+3}$ or between the oscillator $\theta_{N+2}$ with the synchronized group $\theta_0$, determined by the condition Eq.~\eqref{eq_condition} as
\begin{equation}\label{eq_condition_uni}
C = \frac{3}{2}\frac{F(a_{N+2}+a_{N+3})}{F(a_N+a_{N+1}+a_{N+2})},
\end{equation}
where $3/2$ is the ratio of their natural frequency differences. 

When $\mu$ is sufficient small, we have that $F(a)/F(b)=a/b$. From the unimodal property, we have $(a_{N+2}+a_{N+3})/(a_N+a_{N+1}+a_{N+2})<2/3$ hence $C > 1$ in Eq.~\eqref{eq_condition_uni}. The synchronized group $\theta_0$ grows bigger and includes $\theta_{N+2}$. 
The same process also takes place for the oscillators $\theta_i$ with $i<N$ due to the symmetry. 
After such step of synchronization, the central synchronized group includes four oscillators $\theta_{i}$ with $i=N-1,\dots,N+2$. With the increase of $K$ further, the next synchronization condition factor for the oscillators $\theta_{N+3}$ reads
\begin{equation}
C = \frac{5}{2}\frac{F(a_{N+3}+a_{N+4})}{F(a_{N-1}+a_N+a_{N+1}+a_{N+2}+a_{N+3})},
\end{equation}
With similar analysis and the unimodal property we assumed, we find that the oscillator $\theta_{N+3}$ will synchronize with the group $\theta_0$ when $\mu$ is sufficient small. The central synchronized group gets larger. Following this process, with the increase of coupling strength, it is straightforward to show that all the oscillators will be included in this group one by one. This follows directly from the unimodal property and also the linear dependence of $F$ on the weight, as shown in Fig.~\ref{fig_12_oscillator}(a).

On the contrary, when inertia sufficiently increases, such linear dependence of $F$ will be weakened to square root and the self-organization chain will be broken at some points $\theta_{N+n}$. Instead of contributing to the growth of the central synchronized group $\theta_0$, a new cluster will form from the synchronization of $\theta_{N+n}$ and $\theta_{N+n+1}$. Then the new cluster will grow larger with the increase of $K$ to another point where the chain is broken again due to the non-linearity of $F(a)$ and a third synchronized group forms. Continuing this argument we obtain the multi-cluster devil's staircase structure, see Fig.~\ref{fig_12_oscillator}(b). In this way, additional synchronized clusters are formed apart from the central one, forming the oscillatory state as in Fig.~\ref{fig_unimodal}(f).

Secondly, in the bimodal case with sufficient large distance between the two peaks, the system can be approximated as two independent unimodal systems when the coupling strength is small enough. In this case two synchronized clusters will form and grow initially from oscillators with the largest weights, corresponding to the two peaks of the bimodal distribution. The self-organizing takes place independently for each cluster.
For small inertias we have continuous growth from the two peaks of the distribution forming a standing wave, see Fig.~\ref{fig_12_oscillator}(e). For large inertias we have the appearance of several small clusters, see Fig.~\ref{fig_12_oscillator}(f). When the coupling strength $K$ is large enough, these two branches of synchronization processes will merge to one, by creating one large central cluster.

Thirdly, for the case where the two peaks of the bimodal distributions have a large overlap, the synchronization process is more complicated. The two oscillators at each peak are close to each other and the oscillators between them also have relatively large weights. In this case, we need to consider about higher order terms in the synchronization processes, i.e. the synchronization condition of several oscillators. 

Considering the case where $\theta_{N-1}$ and $\theta_{N+2}$ are the two peaks oscillators with the maximum weights, and the other two oscillators $\theta_{N},\theta_{N+1}$ between them have slightly smaller weights. 
The condition of four-oscillator synchronization group $\theta_{N-1},\dots,\theta_{N+2}$ can be estimated by the synchronization condition of the oscillator $\theta_{N+2}$ with the assumed synchronized cluster of $\theta_{N-1},\theta_{N},\theta_{N+1}$. Then the condition factor comparing the appearance of four-oscillator group and the synchronized central group between $\theta_{N+2}$ and $\theta_{N+3}$  reads
\begin{equation}
C = \frac{3}{2}\frac{F(a_{N+2}+a_{N+3})}{F(a_{N-1}+a_N+a_{N+1}+a_{N+2})}.
\end{equation}
From the fact that $a_{N-1}=a_{N+2}>a_{N+3}$ as the maximum weight, if the two central oscillators have relatively large weights $a_N+a_{N+1}>a_{N+3}$, we have $C<1$ and observe the abrupt appearance of four-oscillator synchronization group when $\mu$ is small, as shown in Fig.~\ref{fig_12_oscillator}(c). On the contrary, if such weight is weakened by the inertia effect, the synchronization process will start from the appearance of two clusters of $\theta_{N-2},\theta_{N-1}$ and $\theta_{N+2},\theta_{N+3}$, forming standing wave states as the one for bimodal cases with smaller overlaps, see Fig.~\ref{fig_12_oscillator}(d). 

\section{Discussion}\label{sec/discussion}

Based on the theoretical analysis in Sec.~\ref{sec/time-periodic-mean-field} and the simplified models in Sec.~\ref{sec/three} we conclude that the main effect of inertias is the weakening of the synchronization influence of giant synchronized clusters on the other oscillators, when the system is in the lower branch of hysteresis loops.
As a result, additional synchronized clusters appear besides the giant clusters when $\mu$ is sufficiently large, thus leading to the appearance of oscillatory states or standing waves.

\ack{
  We thank the Center for Information Technology of the University of Groningen for the use of the Peregrine HPC cluster.
J.\ Gao acknowledges support by a China Scholarship Council (CSC) scholarship.}

\section*{References}
\bibliographystyle{unsrt}
\bibliography{papersnew}

\begin{thebibliography}{10}

\bibitem{Arenas2008}
Alex Arenas, Albert D{\'\i}az-Guilera, Jurgen Kurths, Yamir Moreno, and
  Changsong Zhou.
\newblock Synchronization in complex networks.
\newblock {\em Physics Reports}, 469(3):93--153, 2008.

\bibitem{Acebron2005kuramoto}
Juan~A Acebr{\'o}n, Luis~L Bonilla, Conrad J~P{\'e}rez Vicente, F{\'e}lix
  Ritort, and Renato Spigler.
\newblock The kuramoto model: A simple paradigm for synchronization phenomena.
\newblock {\em Reviews of modern physics}, 77(1):137, 2005.

\bibitem{Kuramoto1987}
Yoshiki Kuramoto and Ikuko Nishikawa.
\newblock Statistical macrodynamics of large dynamical systems. case of a phase
  transition in oscillator communities.
\newblock {\em Journal of Statistical Physics}, 49(3):569--605, 1987.

\bibitem{Rodrigues2016}
Francisco~A Rodrigues, Thomas K~DM Peron, Peng Ji, and J{\"u}rgen Kurths.
\newblock The kuramoto model in complex networks.
\newblock {\em Physics Reports}, 610:1--98, 2016.

\bibitem{Ermentrout1991}
Bard Ermentrout.
\newblock An adaptive model for synchrony in the firefly pteroptyx malaccae.
\newblock {\em Journal of Mathematical Biology}, 29(6):571--585, 1991.

\bibitem{Levi1978}
M~Levi, Frank~C Hoppensteadt, and WL~Miranker.
\newblock Dynamics of the josephson junction.
\newblock {\em Quarterly of Applied Mathematics}, 36(2):167--198, 1978.

\bibitem{Watanabe1994}
Shinya Watanabe and Steven~H Strogatz.
\newblock Constants of motion for superconducting josephson arrays.
\newblock {\em Physica D: Nonlinear Phenomena}, 74(3-4):197--253, 1994.

\bibitem{Trees2005}
BR~Trees, V~Saranathan, and D~Stroud.
\newblock Synchronization in disordered josephson junction arrays: Small-world
  connections and the kuramoto model.
\newblock {\em Physical Review E}, 71(1):016215, 2005.

\bibitem{Ikeda2012}
Yuichi Ikeda, Hideaki Aoyama, Yoshi Fujiwara, Hiroshi Iyetomi, Kazuhiko
  Ogimoto, Wataru Souma, and Hiroshi Yoshikawa.
\newblock Coupled oscillator model of the business cycle with fluctuating goods
  markets.
\newblock {\em Progress of Theoretical Physics Supplement}, 194:111--121, 2012.

\bibitem{Sakyte2011}
Edita Sakyte and Minvydas Ragulskis.
\newblock Self-calming of a random network of dendritic neurons.
\newblock {\em Neurocomputing}, 74(18):3912--3920, 2011.

\bibitem{Filatrella2008}
Giovanni Filatrella, Arne~Hejde Nielsen, and Niels~Falsig Pedersen.
\newblock Analysis of a power grid using a kuramoto-like model.
\newblock {\em The European Physical Journal B}, 61(4):485--491, 2008.

\bibitem{Rohden2012}
Martin Rohden, Andreas Sorge, Marc Timme, and Dirk Witthaut.
\newblock Self-organized synchronization in decentralized power grids.
\newblock {\em Physical Review Letters}, 109(6):064101, 2012.

\bibitem{Rohden2014}
Martin Rohden, Andreas Sorge, Dirk Witthaut, and Marc Timme.
\newblock Impact of network topology on synchrony of oscillatory power grids.
\newblock {\em Chaos: An Interdisciplinary Journal of Nonlinear Science},
  24(1):013123, 2014.

\bibitem{Lozano2012}
Sergi Lozano, Lubos Buzna, and Albert D{\'\i}az-Guilera.
\newblock Role of network topology in the synchronization of power systems.
\newblock {\em The European Physical Journal B}, 85(7):1--8, 2012.

\bibitem{Witthaut2012}
Dirk Witthaut and Marc Timme.
\newblock Braess's paradox in oscillator networks, desynchronization and power
  outage.
\newblock {\em New Journal of Physics}, 14(8):083036, 2012.

\bibitem{Menck2013}
Peter~J Menck, Jobst Heitzig, Norbert Marwan, and J{\"u}rgen Kurths.
\newblock How basin stability complements the linear-stability paradigm.
\newblock {\em Nature Physics}, 9(2):89--92, 2013.

\bibitem{Hellmann2016}
Frank Hellmann, Paul Schultz, Carsten Grabow, Jobst Heitzig, and J{\"u}rgen
  Kurths.
\newblock Survivability of deterministic dynamical systems.
\newblock {\em Scientific Reports}, 6, 2016.

\bibitem{Kim2015}
Heetae Kim, Sang~Hoon Lee, and Petter Holme.
\newblock Community consistency determines the stability transition window of
  power-grid nodes.
\newblock {\em New Journal of Physics}, 17(11):113005, 2015.

\bibitem{Gambuzza2017}
Lucia~Valentina Gambuzza, Arturo Buscarino, Luigi Fortuna, Maurizio Porfiri,
  and Mattia Frasca.
\newblock Analysis of dynamical robustness to noise in power grids.
\newblock {\em IEEE Journal on Emerging and Selected Topics in Circuits and
  Systems}, 2017.

\bibitem{Dorfler2013c}
Florian D{\"o}rfler, Michael Chertkov, and Francesco Bullo.
\newblock Synchronization in complex oscillator networks and smart grids.
\newblock {\em Proceedings of the National Academy of Sciences},
  110(6):2005--2010, 2013.

\bibitem{Grzybowski2016}
JMV Grzybowski, EEN Macau, and T~Yoneyama.
\newblock On synchronization in power-grids modelled as networks of
  second-order kuramoto oscillators.
\newblock {\em Chaos: An Interdisciplinary Journal of Nonlinear Science},
  26(11):113113, 2016.

\bibitem{Maizi2016}
Nadia Ma{\"\i}zi, Vincent Krakowski, Edi Assoumou, Vincent Mazauric, and Xiang
  Li.
\newblock Time reconciliation and space agregation to shed light on the
  plausibility of long-term low carbon pathways for power systems.
\newblock In {\em Smart Energy Grid Engineering (SEGE), 2016 IEEE}, pages
  106--110. IEEE, 2016.

\bibitem{Manik2016a}
Debsankha Manik, Martin Rohden, Henrik Ronellenfitsch, Xiaozhu Zhang, Sarah
  Hallerberg, Dirk Witthaut, and Marc Timme.
\newblock Network susceptibilities: Theory and applications.
\newblock {\em Physical Review E}, 95(1):012319, 2017.

\bibitem{Pinto2016}
Rafael~S Pinto and Alberto Saa.
\newblock Synchrony-optimized networks of kuramoto oscillators with inertia.
\newblock {\em Physica A: Statistical Mechanics and its Applications},
  463:77--87, 2016.

\bibitem{Rohden2017}
Martin Rohden, Dirk Witthaut, Marc Timme, and Hildegard Meyer-Ortmanns.
\newblock Curing critical links in oscillator networks as power flow models.
\newblock {\em New Journal of Physics}, 19(1):013002, 2017.

\bibitem{Witthaut2016}
Dirk Witthaut, Martin Rohden, Xiaozhu Zhang, Sarah Hallerberg, and Marc Timme.
\newblock Critical links and nonlocal rerouting in complex supply networks.
\newblock {\em Physical Review Letters}, 116(13):138701, 2016.

\bibitem{Olmi2014}
Simona Olmi, Adrian Navas, Stefano Boccaletti, and Alessandro Torcini.
\newblock Hysteretic transitions in the kuramoto model with inertia.
\newblock {\em Physical Review E}, 90(4):042905, 2014.

\bibitem{Tanaka1997}
Hisa-Aki Tanaka, Allan~J Lichtenberg, and Shin'ichi Oishi.
\newblock First order phase transition resulting from finite inertia in coupled
  oscillator systems.
\newblock {\em Physical Review Letters}, 78(11):2104, 1997.

\bibitem{Tanaka1997a}
Hisa-Aki Tanaka, Allan~J Lichtenberg, and Shin'ichi Oishi.
\newblock Self-synchronization of coupled oscillators with hysteretic
  responses.
\newblock {\em Physica D: Nonlinear Phenomena}, 100(3):279--300, 1997.

\bibitem{Acebron2000synchronization}
JA~Acebr{\'o}n, LL~Bonilla, and R~Spigler.
\newblock Synchronization in populations of globally coupled oscillators with
  inertial effects.
\newblock {\em Physical Review E}, 62(3):3437, 2000.

\bibitem{Barre2016}
Julien Barre and David M{\'e}tivier.
\newblock Bifurcations and singularities for coupled oscillators with inertia
  and frustration.
\newblock {\em Physical review letters}, 117(21):214102, 2016.

\bibitem{Olmi2016}
Simona Olmi and Alessandro Torcini.
\newblock Dynamics of fully coupled rotators with unimodal and bimodal
  frequency distribution.
\newblock In {\em Control of Self-Organizing Nonlinear Systems}, pages 25--45.
  Springer, 2016.

\bibitem{Gao2018}
Jian Gao and Konstantinos Efstathiou.
\newblock Self-consistent method and steady states of second-order oscillators.
\newblock {\em Physical Review E}, 98(4):042201, 2018.

\bibitem{Martens2009}
Erik~Andreas Martens, E~Barreto, SH~Strogatz, E~Ott, P~So, and TM~Antonsen.
\newblock Exact results for the kuramoto model with a bimodal frequency
  distribution.
\newblock {\em Physical Review E}, 79(2):026204, 2009.

\bibitem{Bonilla1998time}
L.~L. Bonilla, C.~J. Vicente~P{\'e}rez, and R.~Spigler.
\newblock Time-periodic phases in populations of nonlinearly coupled
  oscillators with bimodal frequency distributions.
\newblock {\em Physica D: Nonlinear Phenomena}, 113(1):79--97, 1998.

\bibitem{Engelbrecht2012}
Jan~R. Engelbrecht and Renato Mirollo.
\newblock Structure of long-term average frequencies for kuramoto oscillator
  systems.
\newblock {\em Physical Review Letters}, 109(3):034103, 2012.

\bibitem{Strogatz2014}
Steven~H. Strogatz.
\newblock {\em Nonlinear dynamics and chaos: with applications to physics,
  biology, chemistry, and engineering}.
\newblock Westview Press, 2014.

\bibitem{Guckenheimer2013}
John Guckenheimer and Philip~J Holmes.
\newblock {\em Nonlinear Oscillations, Dynamical Systems, and Bifurcations of
  Vector Fields}, volume~42.
\newblock Springer Science \& Business Media, 2013.

\bibitem{Fenichel1971}
N.~Fenichel.
\newblock Persistence and smoothness of invariant manifolds for flows.
\newblock {\em Indiana University Mathematics Journal}, 21(3):193--226, 1971.

\bibitem{Hirsch1970}
M.~W. Hirsch, C.~C. Pugh, and M.~Shub.
\newblock Invariant manifolds.
\newblock {\em Bull. Amer. Math. Soc.}, 76:1015--1019, 1970.

\bibitem{Devaney2003introduction}
Robert~L. Devaney.
\newblock {\em An Introduction to Chaotic Dynamical Systems}.
\newblock Westview Press, 2 edition, 2003.

\bibitem{Acebron1998asymptotic}
JA~Acebr{\'o}n and LL~Bonilla.
\newblock Asymptotic description of transients and synchronized states of
  globally coupled oscillators.
\newblock {\em Physica D: Nonlinear Phenomena}, 114(3-4):296--314, 1998.

\bibitem{Acebron2001bifurcations}
JA~Acebr{\'o}n, A~Perales, and R~Spigler.
\newblock Bifurcations and global stability of synchronized stationary states
  in the kuramoto model for oscillator populations.
\newblock {\em Physical Review E}, 64(1):016218, 2001.

\end{thebibliography}

\end{document}